\documentclass[sigconf]{acmart}

\copyrightyear{2026}
\acmYear{2026}
\setcopyright{cc}
\setcctype{by}
\acmConference[Middleware '26]{27th International Middleware Conference}{December 14--18, 2026}{Tarragona, Spain}
\acmBooktitle{27th International Middleware Conference (Middleware '26), December 14--18, 2026, Tarragona, Spain}
\acmDOI{10.1145/3801927.3858632}
\acmISBN{979-8-4007-2621-7/2026/12}

\usepackage{booktabs}
\usepackage{multirow}
\usepackage{algpseudocode}
\usepackage{algorithm}
\usepackage{amsthm}
\newtheorem*{theorem*}{Theorem 1}
\usepackage{pgfplots}
\usepackage{pgfplotstable}
\usepackage{tikz}
\usepackage{subcaption}
\definecolor{revisiongreen}{RGB}{0,0,0}
\definecolor{revisionpurple}{RGB}{0,0,0}
\definecolor{revisionorange}{RGB}{0,0,0}
\definecolor{revisionred}{RGB}{0,0,0}
\makeatletter
\g@addto@macro\@authorfont{\color{black}}
\g@addto@macro\@affiliationfont{\color{black}}
\makeatother
\usepgfplotslibrary{dateplot,statistics,groupplots}
\usetikzlibrary{patterns}
\pgfplotsset{compat=1.18}

\begin{document}

\title{STLT: A Novel System for Energy-Aware Distributed Inference in the Edge-Cloud-Space Continuum}

\author{Andrija Stanisic}
\email{a.stanisic@dsg.tuwien.ac.at}
\orcid{0009-0007-2120-1142}
\affiliation[obeypunctuation]{%
  \institution{Distributed Systems Group}\\
  \institution{TU Wien}, \country{Austria}
}

\author{Milos Gravara}
\email{m.gravara@dsg.tuwien.ac.at}
\orcid{0009-0006-7986-5243}
\affiliation[obeypunctuation]{%
  \institution{Distributed Systems Group}\\
  \institution{TU Wien}, \country{Austria}
}

\author{Juan Luis Herrera}
\email{jlherrerag@unex.es}
\orcid{0000-0002-2280-2878}
\affiliation{%
  \institution{University of Extremadura}
  \country{Spain}
}

\author{Stefan Nastic}
\orcid{0000-0003-0410-6315}
\email{s.nastic@dsg.tuwien.ac.at}
\affiliation[obeypunctuation]{%
  \institution{Distributed Systems Group}\\
  \institution{TU Wien}, \country{Austria}
}

\renewcommand{\shortauthors}{Stanisic et al.}
\settopmatter{authorsperrow=4}

\begin{abstract}

Low Earth Orbit satellites are pivotal for Earth observation tasks such as wildfire detection, capturing imagery for analysis by Deep Neural Networks (DNNs). On-board analysis reduces downlink volume, but severe energy constraints force a trade-off between computation and data transmission. Splitting the DNN between satellite and ground turns this balance into a design choice, where the split point determines combined computation and downlink energy. Existing approaches typically rely on static split-point decisions that fail to account for dynamic space conditions, or on complex algorithms, historical data, or prediction models that introduce significant overhead. To address these shortcomings, we present STLT, a two-stage system for energy-aware distributed inference between LEO satellites and a ground station. First, an offline analyzer characterizes the DNN architecture to identify a set of favorable split points, consequently reducing the search space. Second, an on-board algorithm performs lightweight energy-aware split selection, dynamically choosing from this pre-computed set based on the current state of the environment. We evaluate STLT on a real-world satellite telemetry dataset, showing that it reduces the split-candidate search space by 30.7\% on average, lowers decision latency by 25.1\% and energy by 27.5\% on average, and delivers up to 20.1\% more inference tasks than a state-of-the-art satellite-ground co-inference baseline while consuming 8.8\% less energy overall.
\end{abstract}



\maketitle

\section{Introduction}
\label{section:introduction}

Low Earth Orbit (LEO) satellites have seen a proliferation in recent years, with approximately 13{,}500 LEO satellites orbiting the Earth in 2026~\cite{10.1109/SEC62691.2024.00028, MARCELINO2025103577, mcdowell_planet4589}. This proliferation is an opportunity to extend computing beyond Earth, as satellites are actively being equipped with commercial off-the-shelf (COTS) hardware, allowing them to perform computation on-board~\cite{arcas2025futurespacebasedhighlyscalable, Martinez_Gost_2025, 10.1145/3636534.3649371}. Motivated by this possibility, researchers have started exploring the benefits and feasibility of integrating satellites into the traditional Edge-Cloud continuum. This culminated in the so-called Edge-Cloud-Space continuum~\cite{10.1109/SEC62691.2024.00028, 11058621}. By its nature, this environment is highly heterogeneous, comprising nodes spanning from powerful cloud centers to constrained edge devices and computing-enabled satellites~\cite{10.1109/SEC62691.2024.00028, 11336648}. Additionally, the environment is highly mobile, with satellites and mobile devices continuously changing the system topology~\cite{MARCELINO2025103577}.

Despite its inherent complexity, the Edge-Cloud-Space continuum serves as a powerful enabler for Earth Observation (EO) applications, such as wildfire detection and disaster management~\cite{10697471, 10.1109/SEC62691.2024.00028, FOOL, inbook}.
Traditional approaches for enabling Edge-Cloud-Space continuum applications mostly follow a "bent pipe" architecture~\cite{FOOL}. This architecture
involves satellites capturing high-resolution imagery, which is then transmitted as raw input to ground stations for analysis~\cite{SLICE, 10680636, FOOL}. However, LEO satellites are constrained devices with limited communication windows~\cite{ZHANG2025115}. Consequently, managing the sheer volume of generated data creates severe system bottlenecks~\cite{FOOL}. To mitigate these problems, recent research focuses on running Deep Neural Network (DNN) inference directly on the satellite~\cite{SLICE, DNNSplit, Martinez_Gost_2025}. The rationale is that DNNs have historically proven to be efficient compression mechanisms because they consume high-dimensional raw data and extract meaningful, low-dimensional information, consequently reducing the transmission data volumes~\cite{7133169}. However, reducing transmitted data by computing more on-board is beneficial only when the communication energy saved outweighs the additional computation energy. On-board execution of DNNs is also complex, as LEO satellites operate under severe energy constraints due to limited battery capacity and intermittent solar charging during eclipse periods~\cite{11015791}. While utilizing a DNN to reduce input dimensions lowers the requirements for storage and transmission, the inference process is often energy-expensive and highly dependent on the neural network's depth~\cite{8335698}. Building a sustainable solution thus requires careful consideration of the trade-offs introduced by on-board computation~\cite{Martinez_Gost_2025}.

To optimize the trade-offs between the on-board computation and the data transmission, many approaches opt for a distributed inference by splitting DNNs between the satellite and the ground station~\cite{9210567, SLICE, Martinez_Gost_2025, 10620909, 10680636}. A split point is the DNN layer at which on-board execution ends and intermediate results are transmitted to the ground station. By intelligently selecting optimal splitting points, these approaches manage the trade-off between the energy consumed for on-board computation and the energy required for propagating intermediate results to the ground~\cite{SLICE, 10.1145/3037697.3037698}. Existing methods opt for the analysis of DNN layers to identify potential split points~\cite{DNNSplit, 10733633, inproceedings}. However, many approaches face limitations when deployed in dynamic satellite environments. {\color{revisiongreen}In particular, extensive layer-level profiling is impractical for on-board execution because its computation and energy costs grow with model depth, while historical data is difficult to collect and maintain under changing orbital conditions~\cite{DNNSplit, 10.1145/3037697.3037698, 8871124}. Even when model analysis is performed offline, static or computationally expensive split-point selection limits practical adaptation to changing orbital conditions~\cite{10733633, inproceedings}.} Unlike terrestrial edge-cloud settings, where connectivity and power supply are relatively stable, satellite-ground inference operates under different constraints~\cite{10.1145/3636534.3649371, 10.1145/3373376.3378473}. Communication windows are intermittent, determined by orbital mechanics rather than network availability~\cite{9210567, 10.1145/3373376.3378473}. Energy budgets fluctuate continuously due to eclipse-driven battery discharge and variable solar harvesting~\cite{MACAMBIRA2022463, 10.1145/3636534.3649371}. Consequently, the feasibility of a given split point changes across tasks within a single orbit, making static selection policies insufficient. Approaches that achieve fine-grained energy optimization often rely on heavyweight scheduling algorithms whose cost makes online execution impractical~\cite{SLICE, 10738397, AOSR}. {\color{revisionorange}Outside communication windows, split selection at the ground station relies on predicted or previously communicated satellite state. On-board split selection, by contrast, can use the current satellite state to determine whether a split point remains feasible.}

{\color{revisiongreen}In this paper, we propose \textbf{S}a\textbf{T}e\textbf{L}li\textbf{T}e (STLT), a two-stage system that enables energy-aware distributed inference between LEO satellites and a ground station. STLT separates model analysis from runtime split selection. In the offline stage, a DNN analyzer characterizes the on-board computation and downlink transmission associated with each split point. STLT models split-point energy as a positive weighted sum of these dimensions. A minimizer of this per-point objective lies on the lower convex hull of the computation-communication cost space, so the analyzer retains the corresponding layers as a compact candidate set. Performing this analysis offline avoids its computational and energy costs on the satellite. In the online stage, a lightweight on-board algorithm uses the precomputed candidate set to perform admission and energy-aware split selection for each inference task according to the current satellite state. Together, the two stages coordinate inference between the satellite and the ground station under LEO energy and communication constraints. STLT does not require historical telemetry traces or learned prediction models, thus reducing the data requirements of the system.} The main contributions of this work include:

\vspace{-0.4cm}
\begin{itemize}    

    \item \textbf{STLT}: {\color{revisiongreen}\textit{A two-stage system for energy-aware distributed inference} between LEO satellites and ground stations. STLT formalizes satellite-ground co-inference under LEO energy and communication constraints and integrates offline DNN analysis with lightweight on-board admission and split selection.} {\color{black}Across five model-specific nominal workloads and 220 five-orbit telemetry segments, STLT delivers 11.4\% more inference tasks than SLICE while reducing total EO energy consumption by 8.8\%.}

    \item \textbf{ANODA}: \textit{\textbf{An} \textbf{O}ffline \textbf{D}NN \textbf{A}nalyzer} {\color{revisiongreen}that determines the candidate split points of a model by retaining the layers on the lower convex hull of the computation-communication cost space. The retained candidates contain the minimum-energy split point for any positive computation and communication cost weights, while feasibility under the orbital energy and communication constraints is resolved on-board at runtime.} An evaluation across five model architectures shows a 30.7\% average reduction in size of the candidate split set compared to state-of-the-art split-candidate pruning. This reduction lowers on-board per-decision latency by 25.1\% and per-decision energy by 27.5\% on average.

    \item \textbf{ANELA}: \textit{\textbf{An} \textbf{E}nergy-aware \textbf{L}ightweight \textbf{A}lgorithm} {\color{revisiongreen}that performs on-board admission and energy-aware split selection for each inference task using the precomputed candidate split points and the current satellite state. ANELA selects the minimum-energy candidate that satisfies the current energy and communication constraints and rejects the task when no candidate is feasible.} 
Evaluation demonstrates lightweight on-board execution, with ANELA averaging $0.0157\,\mathrm{ms}$ latency and $0.056\,\mathrm{mJ}$ energy per decision across the evaluated model configurations, which makes this algorithm suitable for execution on resource-constrained devices.

\end{itemize}

The remainder of this paper is organized as follows. Section~\ref{section:motivation} motivates the work through a motivational EO use case. Section~\ref{section:system_model} presents a formalization of the system model and the optimization objective. Section~\ref{section:stlt} details the STLT system, specifically focusing on the key underlying mechanisms. Experimental results and insights are presented in Section~\ref{section:evaluation}. Finally, Section~\ref{section:related_work} reviews related work, and Section~\ref{section:conclusion} concludes with future research directions.

\section{Motivational Use Case}
\label{section:motivation}

\begin{figure}[h]
    \centering
    \includegraphics[width=0.8\linewidth]{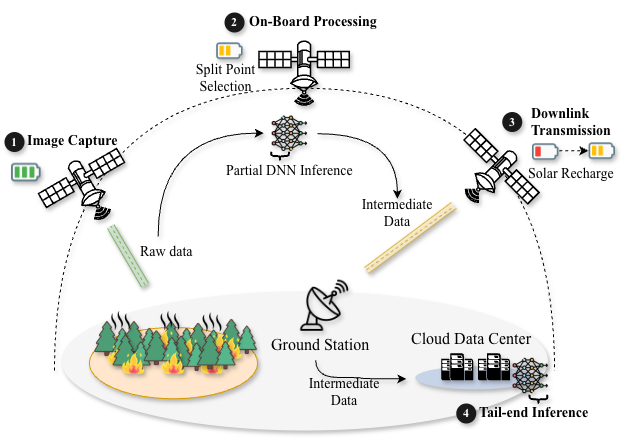}
    \caption{{\color{black}Illustrative wildfire-detection use case, representing distributed inference in the Edge-Cloud-Space continuum.}}
    \label{fig:motivation}
\end{figure}

\vspace{-0.2cm}
Wildfires cause significant human and economic losses worldwide, with annual damages frequently exceeding hundreds of billions of USD~\cite{united2025global}. Early detection is critical to limiting their spread, yet traditional monitoring approaches rely on ground-based sensor networks that require reliable terrestrial network connectivity. Many rural and remote regions, which are disproportionately affected by wildfires, lack such infrastructure~\cite{united2025global}. LEO satellite constellations offer an infrastructure-independent alternative. Equipped with imaging sensors and on-board compute hardware, they provide global coverage and frequent revisit times, enabling continuous EO over affected regions without relying on terrestrial networks~\cite{article}. Consider such a wildfire detection system, as illustrated in Figure~\ref{fig:motivation}. The satellite captures {\color{black}one image approximately every 30 seconds} and processes them using a DNN to identify potential wildfire signatures. {\color{black}The system employs a satellite similar to BUPT-1~\cite{10.1145/3636534.3649371}, equipped with a COTS computing platform whose CPU is equivalent to that of an NVIDIA Jetson Orin Nano for on-board processing.} Inference is performed utilizing {\color{black}SqueezeNet1.1~\cite{iandola2016squeezenet} as the target DNN architecture, applied to 4096$\times$4096 imagery captured by the on-board sensor}. {\color{black}The satellite operates in a 90-minute orbital period, with a 5-minute communication window when positioned above the ground station. During this window, the satellite transmits processed data at 20 Mbps over an X-Band link.} The use case shown in Figure~\ref{fig:motivation} highlights a design choice regarding the processing pipeline. Specifically, the system must determine whether to transmit raw imagery or transmit intermediate results, which result from on-board inference. While on-board computation can reduce transmission volume and improve throughput, it requires dealing with a trade-off between communication and computational energy~\cite{SLICE, 10738397}.
However, in LEO satellite settings this trade-off is fundamentally more constrained than in terrestrial edge-cloud. The available energy budget varies continuously across an orbit due to eclipse periods and solar charging cycles. Similarly, the communication window is a scheduled deadline rather than an on-demand resource. This means that the feasibility of a given split point is not fixed. It depends on the satellite's instantaneous battery state and the accumulated data volume at decision time.


{\color{black}To further illustrate this trade-off, we analyzed real-world telemetry from the BUPT-1 satellite~\cite{10.1145/3636534.3649371}} {\color{black} and measured, on the CPU of an NVIDIA Jetson Orin Nano, the per-layer on-board computation energy at (4096$\times$4096) resolution for SqueezeNet1.1~\cite{iandola2016squeezenet}. For each candidate split point, we execute the network prefix on a single pinned CPU core and measure the dynamic energy per inference using NVIDIA's \texttt{tegrastats}, subtracting the idle baseline and averaging over repeated runs.} This analysis evaluates system throughput, defined as the percentage of captured images successfully transferred to the ground station within one orbital cycle, and per-image energy consumption across two distinct scenarios. In \textit{(i) raw transmission}, the satellite captures images and stores raw image data without performing any on-board inference. During the communication window, accumulated images are transmitted directly to the ground station, where inference is performed. In \textit{(ii) on-board computation}, the satellite processes data on-board up to a specific layer of the DNN during the computation phase. The intermediate output data is stored and subsequently transmitted during the communication phase. To characterize performance across the entire DNN depth, we sample split points at each 10\% interval (decile). {\color{black}The figure shows one layer per decile for readability, whereas the offline model analysis considers every layer, as described in Section~\ref{subsection:offline_analyzer}.}

    

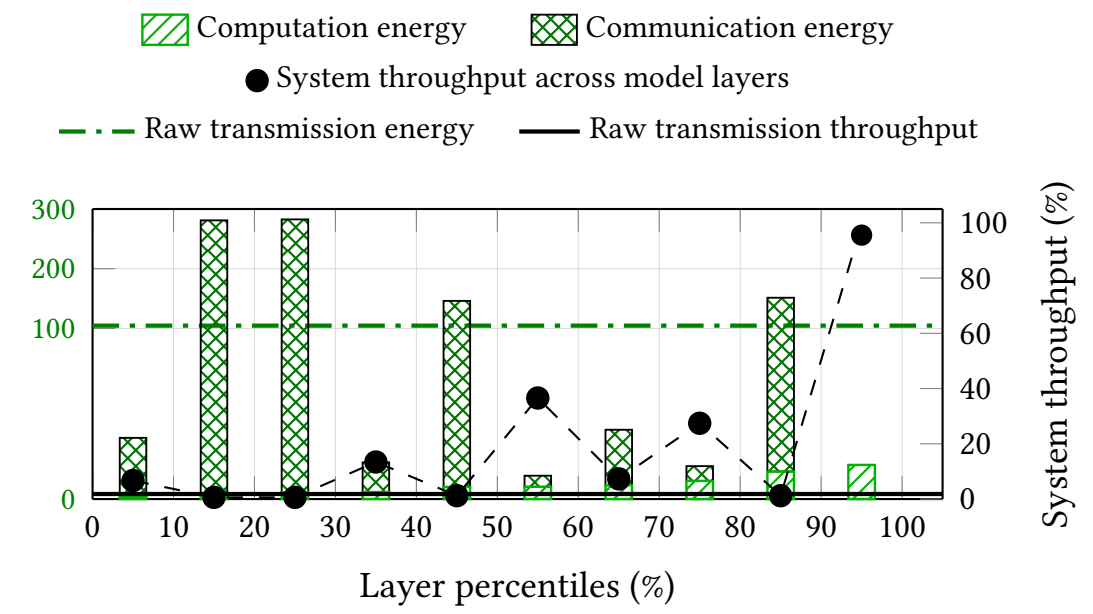
\begin{figure}[h]
    \centering
\begin{tikzpicture}
\begin{axis}[
 width=0.85\columnwidth, height=3.5cm,
 xlabel={Layer percentiles (\%)}, ylabel={Energy (J)},
 ylabel style={green!50!black}, axis y line*=left,
 xmin=0, xmax=105, ymin=0, ymax=170,
 ytick={0,100,135,170}, yticklabels={0,100,200,300}, scaled y ticks=false,
 grid=both, grid style={line width=.1pt, draw=gray!30},
 title style={font=\small\bfseries}, label style={font=\footnotesize},
 tick label style={font=\scriptsize},
 yticklabel style={font=\scriptsize, color=green!50!black},
 xtick={0,10,20,30,40,50,60,70,80,90,100},
 ybar stacked, bar width=5pt, legend style={draw=none}, clip=false, name=ax1
]
\addplot[fill=green!20, draw=green!70!black, line width=.45pt, pattern=north east lines, pattern color=green!70!black] coordinates {(5,1.042438938) (15,2.386376930) (25,3.999102471) (35,4.133233702) (45,7.084119714) (55,7.184324729) (65,8.161323852) (75,10.641398649) (85,16.152675892) (95,20.000000000)};
\addplot[fill=green!5, draw=black, line width=.35pt, pattern=crosshatch, pattern color=green!45!black] coordinates {(5,34.828485120) (15,160.968613331) (25,159.920341730) (35,17.380213760) (45,109.059920714) (55,6.492096000) (65,32.460480000) (75,8.656128000) (85,101.838960670) (95,0.002080000)};
\draw[green!50!black, thick, dash pattern=on 5pt off 2pt on 1pt off 2pt] (axis cs:0,101.641439744) -- (axis cs:105,101.641439744);
\end{axis}
\begin{axis}[
 width=0.85\columnwidth, height=3.5cm,
 ylabel={System throughput (\%)}, ylabel style={black},
 axis y line*=right, axis x line=none,
 xmin=0, xmax=105, ymin=0, ymax=105, ytick={0,20,40,60,80,100},
 tick label style={font=\scriptsize},
 yticklabel style={font=\scriptsize\bfseries, color=black},
 label style={font=\footnotesize}
]
\addplot[black, mark=*, only marks, mark size=2pt] coordinates {(5,6.707317073) (15,0.609756098) (25,0.609756098) (35,13.414634146) (45,1.219512195) (55,36.585365854) (65,7.317073171) (75,27.439024390) (85,1.219512195)};
\fill[black] ([yshift=-2pt]axis cs:95,99.390243902) circle[radius=2pt];
\addplot[black, dashed] coordinates {(5,6.707317073) (15,0.609756098) (25,0.609756098) (35,13.414634146) (45,1.219512195) (55,36.585365854) (65,7.317073171) (75,27.439024390) (85,1.219512195) (95,99.390243902)};
\draw[black, thick, solid] (axis cs:0,1.829268292683) -- (axis cs:105,1.829268292683);
\end{axis}
\node[anchor=south, above=8pt] at (ax1.north) {
\setlength{\tabcolsep}{4pt}\scriptsize
\begin{tabular}{l l}
\multicolumn{2}{c}{
\tikz[baseline={([yshift=-.5ex]current bounding box.center)}]{\draw[fill=green!20, draw=green!70!black, line width=.45pt, pattern=north east lines, pattern color=green!70!black] (0,0) rectangle (3mm,2mm);} Computation energy \hspace{1.5em}
\tikz[baseline={([yshift=-.5ex]current bounding box.center)}]{\draw[fill=green!5, draw=black, line width=.35pt, pattern=crosshatch, pattern color=green!45!black] (0,0) rectangle (3mm,2mm);} Communication energy
} \\[1ex]
\multicolumn{2}{c}{\tikz[baseline={([yshift=-.5ex]current bounding box.center)}]{\draw[fill=black] (0,1mm) circle (2pt);} System throughput across model layers} \\[1ex]
\tikz[baseline={([yshift=-.5ex]current bounding box.center)}]{\draw[green!50!black, thick, dash pattern=on 5pt off 2pt on 1pt off 2pt] (0,1mm) -- (5mm,1mm);} Raw transmission energy &
\tikz[baseline={([yshift=-.5ex]current bounding box.center)}]{\draw[black, thick] (0,1mm) -- (4mm,1mm);} Raw transmission throughput
\end{tabular}};
\end{tikzpicture}
    \caption{{\color{black}Analysis of the computation-communication energy trade-off and its impact on system throughput across different DNN layers.}}
    \label{fig:motivation_throughput}
\end{figure}

Figure~\ref{fig:motivation_throughput} illustrates the computation-communication energy trade-off across model layers. Additionally, it highlights how these energy dynamics directly impact system throughput. {\color{black}The light green diagonally hatched segments represent computation energy consumed while executing DNN inference on the satellite, while the dark green crosshatched segments represent communication energy required to transmit intermediate data to the ground station.} Black dots indicate {\color{black}system throughput} across model layers. For high-resolution imagery (4096×4096), raw transmission achieves {\color{black}1.8\%} throughput and consumes 104.7~J per image. On-board computation enables throughput recovery up to {\color{black}99.4\%} by reducing transmitted data volume. However, energy consumption varies across model layers. Some split points consume significantly more energy than raw transmission while degrading system throughput. For example, split points in the 10--20 percentile range consume up to {\color{black}281.0~J} per image while degrading throughput to {\color{black}0.6\%}, demonstrating both higher energy consumption and worse throughput than raw transmission. {\color{black}These outcomes occur because certain layers expand the intermediate feature map beyond the size of the raw image. Transmitting this larger feature map dominates the per-image energy and costs more than raw transmission.} Conversely, split points in the {\color{black}50--60} percentile range, with approximately {\color{black}13.7~J} of energy consumption per image, achieve {\color{black}36.6\%} {\color{black}system throughput}, {\color{black}20 times} higher than the raw scenario. This wide performance variation demonstrates the importance of intelligent split-point selection. However, selecting an appropriate split point can be inherently challenging due to the discrete nature of computation and communication costs across layers. Prior work has shown that split-point selection problems under time-varying constraints often lead to NP-hard complexity, where evaluating all feasible configurations becomes impractical as model depth and structural complexity increase~\cite{DADS}. This motivates the need for a lightweight automated mechanism to adapt split decisions to dynamic environments.

\vspace{-0.5cm}
\section{STLT System Model}
\label{section:system_model}

In this section, we first present satellite-to-ground co-inference model in Section~\ref{subsection:sat-to-ground}. We then present energy models for on-board computation, data transmission, and solar gains in Section~\ref{section:energy_consumption_model}. Finally, we formulate the optimization problem to minimize total energy consumption across the inference workload in Section~\ref{subsection:problem_definition}.

\subsection{Satellite-to-Ground Co-Inference Model}\label{subsection:sat-to-ground}


Our model focuses on energy characteristics between a single LEO satellite and a dedicated ground station when executing co-inference workloads. STLT optimizes the satellite-to-ground link, which is the dominant energy bottleneck in this setting. Downlink transmission is substantially more energy-expensive than inter-satellite links {\color{revisiongreen}(ISLs)}~\cite{10930516}, so it is where energy optimization yields the largest gains. We therefore do not model {\color{revisiongreen}ISLs} or constellation coordination. Constellation-wide inference over {\color{revisiongreen}ISLs} is a complementary problem revisited as future work in Section~\ref{section:conclusion}.

\vspace{-0.2cm}

\begin{figure}[h]
    \centering
    \includegraphics[width=\linewidth]{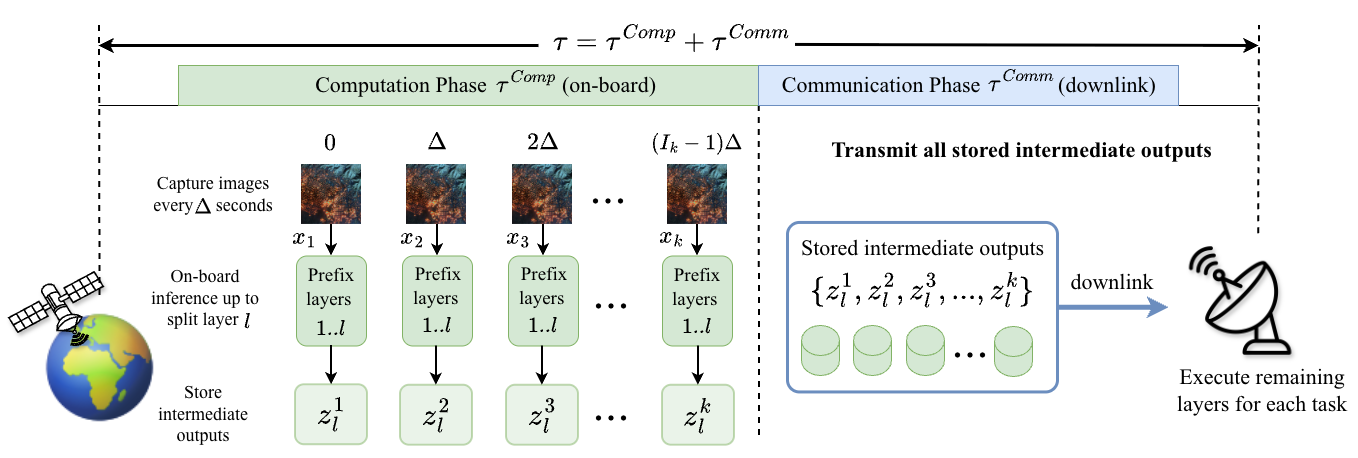}
    \caption{Satellite-to-ground co-inference model.}
    \label{fig:stlt-model}
\end{figure}

\vspace{-0.2cm}

For this model, illustrated in Figure~\ref{fig:stlt-model}, let $\mathcal{T}$ denote the orbital period of the LEO satellite. One orbital period comprises a computation phase and a communication phase. The computation phase period is denoted with $\mathcal{T}^{comp}$, while $\mathcal{T}^{comm}$ denotes the communication phase duration. Consequently, the orbital period can be calculated as the sum of these phases, formally expressed as $\mathcal{T} = \mathcal{T}^{comp} + \mathcal{T}^{comm}$. During computation phase, the satellite captures images and executes DNN workloads. The DNN model architecture is represented as an ordered set of layers $L$, so that $L=\{0, 1, \ldots, |L|-1\}$. The split layer $l \in L$ defines the boundary of on-board execution. It indicates that all DNN layers from the input up to and including layer $l$ are executed on the satellite. {\color{revisiongreen}When $l=0$, no inference is performed on-board and the raw input is transmitted to the ground station.} Conversely, if $l=|L|-1$, complete inference is executed on-board and only the final result is transmitted. We assume that, during the computation phase, the satellite captures pictures at a fixed time step $\Delta$. Therefore, the number of inference tasks $I_k$ for orbit cycle $k$ is determined as $I_k = \lfloor \mathcal{T}^{comp} / \Delta \rfloor$, where the floor function excludes partial intervals when $\mathcal{T}^{comp} \bmod \Delta \neq 0$. {\color{revisiongreen}During the communication phase, the satellite transmits the intermediate data and selected split layer to the ground station.} Upon arrival, the ground station executes the remaining $(|L|-1) - l$ layers of the DNN model. Both the LEO satellite and the ground station are equipped with the same DNN model to satisfy the requirements of satellite-to-ground co-inference model~\cite{SLICE}. The inference workload comprises $I_{total}$ inference tasks. As the satellite captures $I_k$ tasks per orbital cycle, the workload spans $K = \lceil I_{total} / I_k \rceil$ cycles, each offering one communication window whose timing is known a priori from orbital mechanics.
We denote by $N_{rem}$ the number of these communication windows still remaining at a given time.
In time-critical EO applications such as wildfire detection, delivering an inference result beyond its communication window yields diminished operational value~\cite{united2025global}. The satellite therefore retains no downlink backlog across cycles, discarding intermediate data that is not transmitted within its cycle's communication window.

\subsection{Satellite Energy Model}
\label{section:energy_consumption_model}

Energy consumption of a LEO satellite is defined as a sum of the on-board DNN computation $E^{comp}$ and the intermediate data transmission $E^{comm}$. This additive approach can be justified by the sequential nature of the orbital phases. The computation energy is consumed during the computation phase $\mathcal{T}^{comp}$, while transmission energy is consumed during the subsequent communication phase $\mathcal{T}^{comm}$. Since both phases occur within a single orbital cycle $k$, the total energy needed for executing task $i$, denoted as $E_i$, can be formalized with Equation~\ref{energy_base}.
\begin{equation}
    E_i = E^{comp}_i + E^{comm}_i
    \label{energy_base}
\end{equation}

\subsubsection{Computational energy consumption}
For each inference task $i\in \{1,2\dots I_k \}$, the on-board computational energy consumption is formalized in Equation~\ref{computation_energy}.
\begin{equation}
    E^{comp}_i =  \frac{\gamma f^2}{\alpha} \cdot {W(l_i)}
    \label{computation_energy}
\end{equation}
where $W(l_i)$ represents the cumulative computational workload, expressed as the number of Floating Point Operations (FLOPs) required to execute the DNN model up to and including layer $l_i$. Here, $l_i \in L$ denotes the split layer selected for task $i$, representing a key optimization decision. The energy consumption of one CPU cycle is $\gamma f^2$, where $f$ (Hz) represents the CPU frequency of the on-board device and $\gamma$ (J$\cdot$s$^2$) is a constant determined by the switched capacitance~\cite{e75-c_4_371, 9509751}. Finally, $\alpha$ (FLOP) represents the computational throughput of the processor, defined as the number of FLOPs executed per CPU cycle. Following standard CMOS power modeling convention~\cite{e75-c_4_371, 126534}, CPU cycles are treated as dimensionless. 

\subsubsection{Transmission energy consumption}
The energy required to transmit the intermediate data for task $i$ is modeled as the product of data volume and transmission energy efficiency, formalized as:
\begin{equation}
    E_i^{comm} = \beta \cdot D(l_i)
    \label{communication_energy}
\end{equation}
where $D(l_i)$ represents the size of the intermediate output at split layer $l_i$, expressed in bits. The parameter $\beta$ represents the energy cost to propagate a single bit during the communication window of orbit cycle $k$. It is derived from the transmission power $P_{tx}$ and the maximum achievable data rate $R^{max}$:
\begin{equation}
    \beta = \frac{P_{tx}}{R^{max}}
\end{equation}
We assume a fixed transmission power $P_{tx}$. Consequently, the optimization focuses on minimizing communication energy by reducing the transmitted data volume. Similar to prior work~\cite{10185626, 10.1145/3373376.3378473}, we model the upper bound of data rate using the Shannon-Hartley theorem:
\begin{equation}
    R^{max} = B \log_2 \left( 1 + \frac{C}{N} \right)
\end{equation}
where $B$ is the channel bandwidth, $N$ is the received noise power, and $C$ is the received signal power. We model signal power with Equation~\ref{eq:signal_power}, following a well-established literature approach~\cite{9964448, 10.1145/3373376.3378473}.
\begin{equation}
    C = P_{tx} G_t G_r \left( \frac{\lambda}{4\pi S} \right)^2
    \label{eq:signal_power}
\end{equation}
where $G_t$ and $G_r$ are the antenna gains, $\lambda$ represents the carrier wavelength, and $S$ is the slant range between satellite and ground.

\subsubsection{Solar energy harvesting}

The primary energy source of LEO satellites during orbit is energy derived through cell arrays capturing solar radiation~\cite{rauschenbach2012solar}. Approximately 85\% of the energy consumption by LEO satellites is sourced from solar panels~\cite{rauschenbach2012solar, SLICE}, making accurate modeling of this behavior essential. Similar to prior work~\cite{fortescue2011spacecraft}, we model the power generated by the solar cell array, denoted as $P^{solar}$, with Equation~\ref{eq:solar_power}.
\begin{equation}
    P^{solar} = P^{sun} \cdot \cos(\theta)
    \label{eq:solar_power}
\end{equation}
where $P^{sun}$ denotes direct solar power and $\theta$ represents the solar incidence angle, defined as the angle between light and the normal to the solar panel plane. Consequently, for a time period $\tau$ we can estimate energy obtained by the solar cell array as:
\begin{equation}
    E^{solar} = P^{solar} \cdot \tau
    \label{eq:solar_energy}
\end{equation}

\subsection{Energy-Aware Split Point Selection Problem}
\label{subsection:problem_definition}

The objective of split point selection is to minimize the energy consumed by the inference workload under the EO energy budget $E^{budget} = \rho \cdot E^{max}$, where $\rho \in (0,1]$ is the fraction of battery capacity allocated to EO inference. EO inference is a secondary payload sharing the battery with mission-critical subsystems, so $\rho$ is small and the remaining capacity stays reserved for platform operation. Reducing the combined computation and communication energy of each task preserves more of this budget, increasing the number of inference tasks the satellite can deliver within it. We formalize the energy objective in Equation~\ref{optimization_problem}, where each decision variable $l_i$ is the split layer selected for inference task $i$ from the precomputed candidate set $L^*$ (see Section~\ref{subsection:offline_analyzer}). Let $\mathcal{A} \subseteq \{1, \ldots, I_{total}\}$ be the set of tasks that admit a feasible split under the running energy and downlink state, and $\mathcal{A}_k \subseteq \mathcal{A}$ those generated in orbit cycle $k$. The set $\mathcal{A}$ is determined by feasibility rather than chosen to optimize the objective, and tasks with no feasible split are rejected on-board by ANELA. Equation~\ref{optimization_problem} then selects the energy-minimizing split $l_i \in L^*$ for each task in $\mathcal{A}$, bounding total workload energy by $E^{budget}$ and the intermediate data in each communication window by the maximum transferable data volume $D^{max} = \mathcal{T}^{comm} \cdot R^{max}$.


\begin{equation}
    \begin{aligned}
        \min_{\{l_i\}_{i \in \mathcal{A}}} \quad
        & \sum_{i \in \mathcal{A}}
        \left(
        \frac{\gamma f^2}{\alpha} \cdot {W(l_i)} + \beta \cdot D(l_i)
        \right) \\
        \text{s.t.} \quad
        & \sum_{i \in \mathcal{A}} E_i(l_i) \leq E^{budget} \\
        & \sum_{i \in \mathcal{A}_k} D(l_i) \leq D^{max}, \quad \forall k \in \{1, \ldots, K\} \\
        & l_i \in L^{*}, \quad \forall i \in \mathcal{A} \\
        & \alpha, \beta, f, \gamma > 0
    \end{aligned}
    \label{optimization_problem}
\end{equation}

Equation~\ref{optimization_problem} defines a workload-level joint optimization problem. In a constellation deployment, this problem can naturally be distributed across satellites, since each satellite operates with its own local state. STLT therefore addresses split point selection locally on each satellite. Within one satellite, task decisions remain coupled by the available energy budget and the finite downlink capacity of each communication window. An exact solution would require a fixed planning horizon, a specified task sequence, and a predicted resource trajectory for the workload. Such optimization is suitable for offline planning, but is not appropriate for repeated on-board execution before every inference task~\cite{SLICE}. STLT therefore performs lightweight on-board split point selection over $L^*$. This selection enforces the energy and data-volume constraints over running state, but does not claim global optimality with respect to Equation~\ref{optimization_problem}. The complete on-board selection algorithm is described in Section~\ref{section:algorithm}.

\section{STLT System Overview}
\label{section:stlt}

{\color{revisiongreen}STLT coordinates energy-aware distributed DNN inference between a LEO satellite and a ground station through a two-stage design that combines offline model analysis with lightweight on-board admission and split selection. These stages are implemented by ANODA and ANELA, respectively.} 

\vspace{-0.3cm}
\begin{figure}[h]
    \centering
    \includegraphics[width=0.9\linewidth]{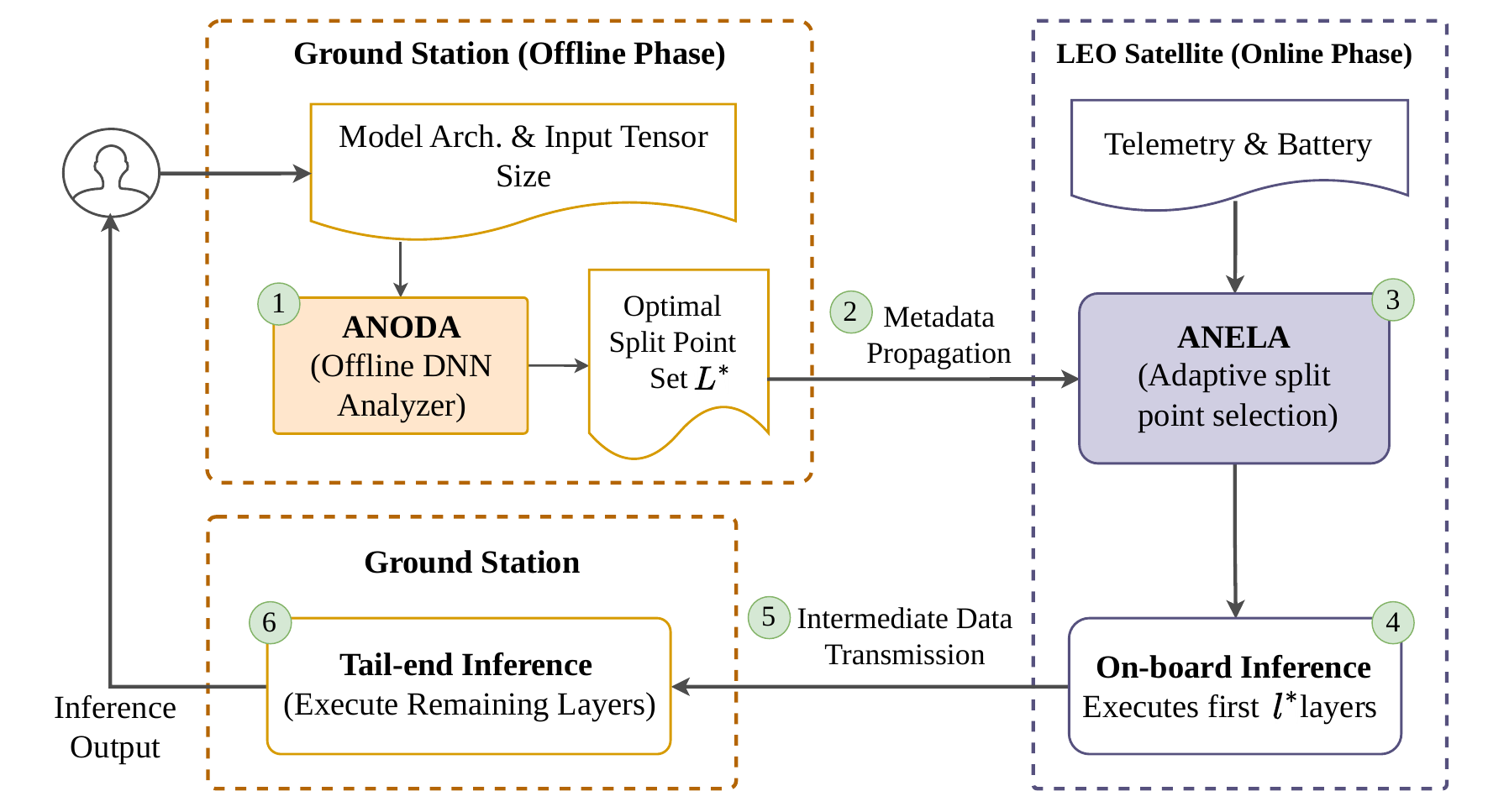}
    \caption{STLT overview depicting distributed inference between LEO satellite and ground station with ANODA, ANELA, and DNN execution stages.}
    \label{fig:stlt-workflow}
\end{figure}

\vspace{-0.2cm}
Figure~\ref{fig:stlt-workflow} illustrates the workflow. In Step~1, the user of STLT defines the inference task by providing a pretrained model and input tensor size. This data is propagated to ANODA, which processes model architecture to compute the convex-hull candidate split set. In Step~2, this set, alongside the provided model, gets propagated to the LEO satellite. During Step~3, for each inference task, ANELA returns either a feasible energy-minimizing split point $l^*$ or a rejection, based on metadata and current satellite state. Before committing on-board computation, ANELA also verifies that the selected candidate preserves enough expected energy for the upcoming communication window. In Step~4, when ANELA returns a split point $l^*$, the satellite executes the first $l = l^*$ layers on the captured imagery and stores the intermediate data. If ANELA returns a rejection, the task is dropped and no on-board computation is performed. {\color{revisiongreen}In Step~5, upon reaching the communication window, the satellite transmits accumulated intermediate data and metadata identifying the inference task, model, selected split point $l^*$, and tensor representation to the ground station. In Step~6, the ground station uses this metadata and the shared model to interpret the data and resume inference from the layer after $l^*$.}
Finally, the generated inference output is made accessible to the end user. This two-stage design directly reflects the operational constraints of LEO satellites. ANODA executes offline on the ground where computational resources are unconstrained, deferring expensive analysis to pre-deployment. ANELA then operates within the satellite's strict energy budget, relying on the precomputed candidate set to keep each decision lightweight and executable onboard.

\subsection{ANODA: Offline DNN Analyzer}
\label{subsection:offline_analyzer}
The primary objective of ANODA is to reduce the search space for the on-board algorithm by identifying a subset of convex-hull candidate split points $L^*$, where $L^* \subseteq L$. By restricting the decision space to this subset, the computational overhead of the on-board algorithm can be significantly reduced. This reduction becomes critical given that the satellite operates under strict energy constraints. The analyzer executes offline on the ground, prior to the initialization of the EO task, and is assumed to have sufficient resources. The analyzer operates in two distinct phases, that is, the $i)$ layer profiling and $ii)$ convex-hull candidate layer selection phase.

\subsubsection{Layer Profiling Phase}
The first step is to characterize the properties of the target DNN architecture. We use automated profiling frameworks to extract layer-wise metadata via static analysis of the model computational graph. Specifically, for each layer $l \in L$ we quantify two metrics. The cumulative computational workload $W(l)$ required to execute the model up to layer $l$, and the intermediate output data size $D(l)$ produced by layer $l$. The profiling process uses an input tensor size specified by the user, ensuring that the extracted metadata accurately reflects the expected operational conditions. The output is a discrete set of operating points \( \mathcal{M} \) embedded in a two-dimensional cost space, formally defined as:
\begin{equation}
    \mathcal{M} = \{(W(l), D(l)) \mid l \in L \}
    \label{discrete_set_m}
\end{equation}
In this representation, the x-axis corresponds to the cumulative computational effort, while the y-axis captures the communication volume for transmitting intermediate data to the ground station. The set $\mathcal{M}$ serves as the input to the subsequent selection phase.

\subsubsection{Convex-Hull Candidate Layer Selection}

The core task of ANODA is to filter the set $\mathcal{M}$ to retain only those layers that can theoretically minimize the total energy consumption under varying orbital conditions. Recall that the total energy is modeled as a linear combination of computation and communication costs. Geometrically, minimizing this linear objective function corresponds to finding a tangent point on the lower boundary of the point set $\mathcal{M}$. {\color{black}This property is a direct consequence of the standard convex-hull characterization of linear optimization, which ensures that at least one minimizer lies among the vertices of the lower convex hull of $\mathcal{M}$~\cite{boyd2004convex}.} The objective function defined in Equation~\ref{optimization_problem} is a linear function of this form, where: $x = W(l)$, $y = D(l)$, $a = (\gamma f^2 / \alpha)$, and $b = \beta$. Note that both coefficients $a,b$ are strictly positive. Consequently, any layer $l$ whose corresponding point $(W(l), D(l))$ lies in the interior of the hull or on a concave section of the boundary is suboptimal. Such points represent inefficient trade-offs where the marginal reduction in transmission volume does not justify the marginal increase in computational effort. ANODA discards these points, ensuring that additional computational energy is invested only if the resulting reduction in transmission data is sufficiently steep to yield a net energy saving. We apply the well-established Andrew's Monotone Chain Algorithm~\cite{Andrew1979AnotherEA} to extract the lower convex hull of model layers. {\color{revisiongreen}This process yields the candidate set $L^*$, which serves as an input to the on-board algorithm. The set retains the minimum-energy split point of the per-point linear objective $f(W,D)$ for any valid energy coefficient pair $(a,b)$.} ANELA checks data-capacity and communication-window energy feasibility online when selecting $l^* \in L^*$ under the orbital constraints of Equation~\ref{optimization_problem}.


\subsection{ANELA: Energy-Aware Selection Algorithm}
\label{section:algorithm}

{\color{revisionorange}Once ANODA embeds the fixed candidate set as metadata, ANELA performs on-board admission and split selection using the current satellite state, as candidate feasibility can change between task arrivals.} ANELA is a greedy per-task online approximation of Equation~\ref{optimization_problem}, designed to fit the resource constraints of onboard execution. At each invocation, ANELA minimizes $E^{comp}(l) + E^{comm}(l)$ over the candidate split points that satisfy its per-task feasibility checks. The workload-level constraints of Equation~\ref{optimization_problem} are correspondingly adapted into per-task feasibility checks over the running satellite state. For each inference task $i$ in orbit cycle $k$, the algorithm returns either a feasible split point $l^*$ from the precomputed candidate set $L^*$ that minimizes total energy consumption, or a rejection when no candidate is feasible under the running state. {\color{revisionorange}At each invocation, $t$ denotes the current position in the computation phase $\mathcal{T}^{comp}$.} Time is reset to zero at the start of each orbital cycle, defined as the end of the previous communication window. At execution time, the algorithm refreshes the remaining EO budget through the state read $\text{queryRemainingBudget}()$. This call returns the remaining EO inference budget $E^{EO}_{rem}$, the persisted allocation reduced by the energy of prior admitted tasks and contracted by any unexpected draw on the shared battery that live telemetry reveals. {\color{black} The remaining budget excludes both energy already spent and energy reserved for queued transmissions.} Reasoning over this live budget enables adaptation to the continuously changing conditions. The algorithm evaluates candidate split points with linear complexity w.r.t. the number of candidate split points identified, making it suitable for resource-constrained satellites. The first filter enforces budget feasibility by verifying that the remaining EO budget exceeds the computational requirement of each candidate $l \in L^{*}$, formally expressed as:
\begin{equation}
   E^{EO}_{rem} \geq \frac{\gamma f^2}{\alpha} \cdot W(l)
\end{equation}
where the right-hand side represents the energy required to execute the DNN up to layer $l$ according to the computational energy model defined in Equation~\ref{computation_energy}. Candidate points that violate this constraint are eliminated from consideration, ensuring on-board computation stays within the remaining EO budget.

The second filter enforces an upper bound on transmitted data volume to prevent exceeding communication channel capacity. Consequently, the remaining candidates must satisfy the following:
\begin{equation}
    l\in L_2^* \Longleftrightarrow D_{sum} + D(l) \leq D^{max}
    \label{eq:d_sum}
\end{equation}
Here, $D_{sum}$ represents the cumulative data volume of split points already committed within the current orbit cycle, $D(l)$ is the data volume contributed by the candidate layer $l$ under evaluation, and $D^{max}$ is the maximum transferable data volume during the communication phase. Equation~\ref{eq:d_sum} corresponds to the data transmission limit constraint defined within the optimization problem in Equation~\ref{optimization_problem}. The role of this constraint is ensuring that admitted intermediate data does not exceed the communication-window capacity. Without this constraint, the system could accumulate data beyond downlink capacity, wasting limited energy resources. The third filter preserves energy for the upcoming communication window. Satellites operate under energy constraints where computation consumes battery capacity that will be needed later for data transmission. The algorithm must therefore verify that each candidate leaves enough expected energy to support the upcoming communication window. To achieve this, we project the remaining EO budget forward to the beginning of the communication window based on predictable orbital characteristics. This estimation accounts for the remaining budget, the EO share of solar gains during the remaining computation period, and the energy consumed by the current task. Formally, the expected EO budget is:
\begin{equation}
    \footnotesize
    E^{expected}_k = \min \{E^{budget}, {\color{black}E^{EO\prime}_{rem}} + \rho(\mathcal{T}^{comp} - t) \cdot P^{solar} - E^{comp}(l)\}
    \label{eq:expected_min}
\end{equation}
{\color{black}where $E^{EO\prime}_{rem}=E^{EO}_{rem}+\beta D_{sum}$ denotes the remaining EO budget including energy reserved for queued transmissions.} The term $(\mathcal{T}^{comp} - t)$ represents remaining time until the communication window begins. The product $\rho(\mathcal{T}^{comp} - t) \cdot P^{solar}$ is therefore the EO share of solar energy harvested during this period, defined in Equation~\ref{eq:solar_energy}. The final term subtracts the energy consumed to execute the DNN up to split layer $l$, defined in Equation~\ref{computation_energy}. The $\min$ operation ensures the expected EO budget does not exceed the allocation $E^{budget}$. Having estimated the expected EO budget at the start of transmission, the algorithm verifies that the candidate leaves enough expected energy for the upcoming communication window. This energy reservation is formalized in Equation~\ref{equation:expected_energy}.
\begin{equation}
    E^{expected}_k + \rho \mathcal{T}^{comm} \cdot P^{solar} \geq \beta (D_{sum} + D(l))
    \label{equation:expected_energy}
\end{equation}
The left-hand side is the energy expected to be available during the communication phase, while the right-hand side is the energy required to transmit the cumulative intermediate data $D_{sum} + D(l)$ stored for the upcoming communication window. Candidate points that violate this condition are eliminated from consideration. The algorithm queries satellite telemetry before each inference task, refreshing the current state at each execution. This enables the algorithm to adjust split point selection based on continuously changing orbital conditions rather than relying on static plans. Finally, a fourth filter admits a candidate only while the remaining EO budget covers its full task energy $E(l) = E^{comp}(l) + E^{comm}(l)$. This keeps the cumulative inference energy within $E^{budget}$ as required by the optimization problem in Equation~\ref{optimization_problem}.
Beyond per-task feasibility, ANELA reserves EO budget across the remaining communication windows. The $N_{rem}$ windows can downlink at most $\beta D^{max} N_{rem}$ of energy. When the remaining budget $E^{EO}_{rem}$ falls to or below this amount, energy rather than downlink capacity limits delivery, and ANELA restricts evaluation to the minimum-energy split point. Otherwise it evaluates the full candidate set $L^*$. From the feasible points, the algorithm selects the one minimizing the objective in Equation~\ref{optimization_problem}, or drops the task when none qualify. The energy of the selected task is charged against the remaining budget.

Algorithm~\ref{alg:stlt_onboard} realizes this selection on-board. ANELA takes the running state $D_{sum}$, the remaining window count $N_{rem}$, and the remaining budget $E^{EO}_{rem}$ as input (line~\ref{alg:dsum_input}). It refreshes $E^{EO}_{rem}$ from telemetry through $\text{queryRemainingBudget}()$ (line~\ref{alg:query}). $D_{sum}$ records the intermediate data stored in the current orbit cycle and resets each cycle. The budget $E^{EO}_{rem}$ starts at $E^{budget}$ and persists across cycles. 
The energy reservation (line~\ref{alg:reserve}) restricts evaluation to the minimum-energy split point when the budget cannot outlast the remaining downlink windows, and otherwise keeps the full set $L^*$. For each candidate (line~\ref{alg:loop}), ANELA computes the computational energy (line~\ref{alg:ecomp}) and applies the four feasibility filters derived above. They enforce budget feasibility against the remaining budget (line~\ref{alg:c1}), channel capacity under $D^{max}$ (line~\ref{alg:c2}), enough projected energy for the upcoming communication window (line~\ref{alg:c3}), and a full-task budget gate that keeps the cumulative inference energy within $E^{budget}$ (line~\ref{alg:c4}). ANELA then selects the feasible split that minimizes the objective (lines~\ref{alg:min_check}--\ref{alg:update}). If no split is feasible, ANELA rejects the task and leaves $D_{sum}$ and $E^{EO}_{rem}$ unchanged (line~\ref{alg:reject}), so the satellite performs neither computation nor transmission and preserves energy for later tasks. Otherwise it commits the split, updates $D_{sum}$ by the output data size of the selected layer, and reduces $E^{EO}_{rem}$ by the task energy $E(l^*)$ (lines~\ref{alg:dsum_update}--\ref{alg:return}).

\begin{algorithm}[h]
\footnotesize
\caption{ANELA}
\begin{algorithmic}[1]
\State \textbf{Input:} $L^{*}$, $\mathcal{T}^{comp}$, $\mathcal{T}^{comm}$, $D^{max}$, $E^{budget}$, $P^{solar}$, $\rho$, $f$, $\beta$, $\alpha$, $\gamma$, $t$, $D_{sum}$, $N_{rem}$, $E^{EO}_{rem}$ \label{alg:dsum_input}
\State \textbf{Output:} $l^*$, updated $D_{sum}$, $E^{EO}_{rem}$
\State $E^{EO}_{rem} \leftarrow \text{queryRemainingBudget}()$ \label{alg:query}
\State $l^*\leftarrow\emptyset$ \label{alg:init}
\State \textbf{if} $E^{EO}_{rem} \leq \beta D^{max} N_{rem}$ \textbf{then} \label{alg:reserve}
\State \ \ \ \ $L_{eval} \leftarrow \{\arg\min_{l \in L^{*}} (\frac{\gamma f^2}{\alpha} W(l) + \beta D(l))\}$
\State \textbf{else}
\State \ \ \ \ $L_{eval} \leftarrow L^{*}$
\For{$l \in L_{eval}$} \label{alg:loop}
    \State $E^{comp}(l) \leftarrow \frac{\gamma f^2}{\alpha}W(l)$ \label{alg:ecomp}
    \State \textbf{if} $E^{EO}_{rem} < E^{comp}(l)$ \textbf{then} \label{alg:c1}
    \State \ \ \ \ \textbf{continue;}

    \State \textbf{if} $D_{sum} + D(l) > D^{max}$ \textbf{then} \label{alg:c2}
    \State \ \ \ \ \textbf{continue;}

    \State $E^{expected} \leftarrow \min \{E^{budget}, {\color{black}E^{EO\prime}_{rem}} + \rho(\mathcal{T}^{comp} - t) \cdot P^{solar} - E^{comp}(l)\}$ \label{alg:expected}
    \State $E^{comm}(l) \leftarrow \beta D(l)$ \label{alg:ecomm_val}

    \State \textbf{if} $E^{expected} + \rho \mathcal{T}^{comm} \cdot P^{solar} < \beta (D_{sum} + D(l))$ \textbf{then} \label{alg:c3}
    \State \ \ \ \ \textbf{continue;}

    \State $E(l) \leftarrow E^{comp}(l)+E^{comm}(l)$ \label{alg:total_e}
    \State \textbf{if} $E^{EO}_{rem} < E(l)$ \textbf{then} \label{alg:c4}
    \State \ \ \ \ \textbf{continue;}
    \State \textbf{if} {$l^* = \emptyset \text{ or } E(l) < E({l^{*}})$} \textbf{then} \label{alg:min_check}
    \State \ \ \ \ \ $l^* \leftarrow l$ \label{alg:update}
\EndFor
\State \textbf{if} $l^* = \emptyset$ \textbf{return} $(\textit{reject}, D_{sum}, E^{EO}_{rem})$ \label{alg:reject}
\State $D_{sum} \leftarrow D_{sum} + D(l^{*})$ \label{alg:dsum_update}
\State $E^{EO}_{rem} \leftarrow E^{EO}_{rem} - E(l^{*})$ \label{alg:budget_update}
\State \Return $(l^{*}, D_{sum}, E^{EO}_{rem})$ \label{alg:return}
\end{algorithmic}
\label{alg:stlt_onboard}
\end{algorithm}

\vspace{-0.3cm}
\section{Evaluation}
\label{section:evaluation}

This section presents a series of experiments to evaluate STLT and analyze its results. Section~\ref{subsection:evaluation_setup} details the experiments carried out and their objectives. Section~\ref{subsection:evaluation_results} analyzes the results.

\subsection{Experimental setup}\label{subsection:evaluation_setup}

The STLT system is evaluated by analyzing its two primary modules, that is, the ANODA analyzer and the ANELA selection algorithm. The experimental methodology examines both the standalone performance of each module and their combined effects, with comparisons against established baselines. This evaluation strategy enables validation of the STLT effectiveness in both search space reduction and energy optimization under dynamic conditions. 

To evaluate the standalone performance of ANODA, we compare it against DNNSplit~\cite{DNNSplit}. DNNSplit is a partitioning strategy designed for multi-tier edge-cloud environments that identifies a subset of split points to minimize latency and cost. Although DNNSplit was not originally evaluated for satellite environments, the authors present their analyzer as a general solution, making it a representative baseline for evaluating ANODA. For a fair comparison, we implement the DNNSplit analyzer logic based on the equations proposed in their work. 

To evaluate ANELA and the end-to-end behavior of STLT, we use two groups of baselines that target distinct stages of the system. The first group consists of naive per-task selection baselines for ANELA. As we are not aware of lightweight online policies designed for dynamic satellite-ground co-inference, we construct three baselines capturing fundamental computation-communication trade-offs, each selecting a split layer from the candidate set $L^*$ without adapting its selection rule to resource availability. Traditional on-ground computing (TOGC) represents the bent-pipe architecture, selecting the split with minimal on-board computation and thus maximizing data forwarded to the ground. Traditional on-board computing (TOBC) is the opposite extreme, selecting the split with minimal communication energy to minimize transmission. Random selection (RAND) selects a split uniformly at random for each task as a sanity check that the optimization problem is non-trivial. {\color{black}The second group consists of SLICE~\cite{SLICE}, a state-of-the-art satellite-ground co-inference framework for LEO systems that computes a layer-wise schedule offline. We reimplement SLICE following the objective and equations presented in its paper and compute one complete-workload plan over all model layers for each experimental configuration. Because this plan remains fixed, the comparison evaluates whether ANELA's per-task adaptation improves delivery under changing satellite conditions at comparable energy cost.}

The evaluation framework for ANODA is implemented leveraging the PyTorch ecosystem for model management and inference simulation~\cite{10.5555/3454287.3455008}. To perform a granular model layer analysis, we employ the \textit{torch.fx} library~\cite{pytorch_fx}. We extract the cumulative floating-point operations $W(l)$ and the size of output data $D(l)$ for each DNN layer. All models are analyzed using a fixed input tensor shape $(3, 4096, 4096)$, matching the high-resolution imagery considered in our satellite scenario. {\color{black}Raw inputs and intermediate tensors are represented in 32-bit floating-point format.} To ensure the generality of the STLT system, we selected a diverse set of DNN architectures.
{\color{black} The considered models cover five architecture families, namely the lightweight SqueezeNet1.1~\cite{iandola2016squeezenet} with 67 layers, the vision transformer Swin-V2-T~\cite{liu2022swin} with 330 layers, the compound-scaled EfficientNet-B4~\cite{tan2019efficientnet} with 503 layers, the residual ResNet50~\cite{he2016deep} with 176 layers, and the densely connected DenseNet169~\cite{huang2017densely} with 600 layers. The number of layers of each model is obtained through the \textit{torch.fx} library.}

{\color{black}

The evaluation framework for ANELA is implemented in Python, following the logic detailed in Algorithm~\ref{alg:stlt_onboard}. To measure its on-board control overhead, we execute ANELA on the Arm Cortex-A78AE CPU of an NVIDIA Jetson Orin Nano, representative of modern COTS hardware used for satellite computing~\cite{10.1145/3636534.3649371}, and measure controller energy consumption using NVIDIA’s \texttt{tegrastats} utility. These measurements concern controller execution, while the inference energy is estimated from cumulative FLOPs using the same analytical CPU model across all evaluated policies. To evaluate policy performance under time-varying orbital conditions, we use a trace-driven simulation based on the complete 125-day BUPT-1 telemetry dataset~\cite{10.1145/3636534.3649371}. The dataset contains gaps between consecutive telemetry records, during which changes in battery state are unobserved. Simulating across these gaps would require assumptions about the missing battery dynamics. We therefore split the dataset at each gap and divide every uninterrupted portion into complete, non-overlapping 27,000~s segments, discarding incomplete remainders. This yields 220 segments, each spanning five orbital cycles of 5,100~s of computation followed by a 300~s communication window. Simulation state and policy accounting are initialized independently for each segment, and the telemetry is replayed sequentially to update the modeled battery energy every second from recorded solar input and platform consumption. Across segments, the sunlit fraction ranges from 56.2\% to 70.3\%, while the net battery-energy change ranges from a 365.1~kJ decrease to a 141.2~kJ increase. These diverse conditions allow us to assess whether ANELA adapts its admission and split-point decisions to both energy-depleting and energy-replenishing orbital conditions.


The considered model architectures differ substantially in computation demand and intermediate-data size, so a common task-generation interval ($\Delta$) would impose different resource pressure across models. At sufficiently short intervals, SLICE cannot construct a feasible complete-workload plan, preventing comparison against a feasible nominal baseline. We therefore progressively increase $\Delta$ for each model until SLICE first finds a feasible plan under the nominal configuration, and adopt this interval as nominal. The resulting intervals are 31, 365, 119, 319, and 269~s for SqueezeNet1.1, Swin-V2-T, EfficientNet-B4, ResNet50, and DenseNet169, respectively, yielding 820, 65, 210, 75, and 90 tasks per five-orbit segment under the simulated task-arrival schedule. These SLICE plans consume 93.3--100.0\% of the nominal EO budget. All workloads use the same 220 segments, task arrivals, and resource settings across policies for each model. Table~\ref{table:evaluation_parameters} lists the nominal parameters.}

\vspace{-0.2cm}

\begin{table}[h]
\footnotesize
\centering
\caption{{\color{black} Nominal evaluation parameters.}}
\begin{tabular}{p{0.30\columnwidth}p{0.33\columnwidth}p{0.20\columnwidth}}
\toprule
Parameter & Value & Unit \\
\midrule
$\gamma$ & $1.67 \cdot 10^{-28}$ & J$\cdot$s$^2$ \\
$f$ & $1.5 \cdot 10^{9}$ & Hz \\
$\alpha$ & 6.0 & FLOP/cycle \\
$\beta$ & $6.5 \cdot 10^{-8}$ & J/bit \\
$\Delta$ & [31, 365, 119, 319, 269] & s \\
$T^{comp}$ & 5100 & s \\
$T^{comm}$ & 300 & s \\
$R^{max}$ & 20 & Mbps \\
$\rho$ & 0.016 & --- \\
$P^{sun}$ & 0--50 & W \\
$E^{max}$ & $8.28 \cdot 10^{5}$ & J \\


\bottomrule
\end{tabular}
\label{table:evaluation_parameters}
\end{table}


{\color{black} At the beginning of each segment, EO inference is allocated $E^{budget}=\rho \cdot E^{max}=13{,}248$~J, corresponding to $\rho=0.016$ of the total battery capacity. During trace replay, changes in total battery energy produce proportional changes in this allocation, with charging restoring it up to $E^{budget}$. Thus, the EO allocation reflects the solar input and shared platform consumption captured by the telemetry. At each task arrival, ANELA reads the current on-board battery state and accounts for energy committed to prior admissions when making admission and split-point decisions.}


The evaluation assesses STLT across three dimensions. First, ANODA is compared with DNNSplit to quantify candidate-set reduction and its impact on per-decision selector cost. {\color{black}Second, trace-driven simulations across five model architectures evaluate STLT end-to-end task delivery, admission and rejection, total and wasted EO energy, and sensitivity to the EO-inference budget fraction and maximum downlink rate against the established baselines.} {\color{black}Third, on-board decision overhead is evaluated on the Jetson Orin Nano CPU to characterize how ANELA scales with the candidate-set size on satellite hardware.} {\color{black}To ensure reproducibility, the complete STLT implementation and evaluation scripts are made publicly available\footnote{\textcolor{black}{\url{https://github.com/polaris-slo-cloud/STLT}}\label{fn:artifact}}.}


\subsection{Result analysis}
\label{subsection:evaluation_results}

{\color{black}The results presented in this section show how STLT's two-stage design translates into practical gains under orbital conditions. The analysis starts from the split points identified offline and then follows their effect through the online stage across the 220 five-orbit telemetry segments. The final part examines whether the resulting per-task decisions remain affordable for on-board execution.}

\subsubsection{ANODA results}
\label{subsubsection:offline_analyzer_evaluation}

The first analysis compares the split points identified by ANODA and DNNSplit. Table~\ref{table:identified_points_summary} summarizes the results across all five architectures. The two approaches apply different selection criteria. DNNSplit marks a layer as a candidate if it reduces intermediate data volume. ANODA applies the convex-hull criterion, retaining only layers that can minimize total energy for some computation-communication weight pair. Across evaluated model architectures, the ANODA candidate set is never larger than the one identified by DNNSplit.

\begin{table}[h]
\footnotesize
\centering
\caption{Identified split layers by DNNSplit and ANODA.}
\label{tab:optimal_points_comparison}
\begin{tabular}{lcc}
\toprule
\textbf{Model} & \textbf{DNNSplit} & \textbf{ANODA} \\
\midrule
SqueezeNet1.1 & [0,4,19,34,65] & [0,4,19,34,65] \\
Swin-V2-T & [0,112,271,327] & [0,112,327] \\
EfficientNet-B4 & [0,40,102,164,352,499,502] & [0,40,102,164,502] \\
ResNet-50 & [0,151,173,175] & [0,175] \\
DenseNet-169 & [0,51,140,369,597,599] & [0,140,599] \\
\bottomrule
\end{tabular}
\label{table:identified_points_summary}
\end{table}


We illustrate the pruning rationale using \texttt{swin\_v2\_t} as a representative case with clear differences in split-point marginal utility, restricting detailed layer-level analysis to this model due to space constraints. The pruning decision is driven by the relationship between cumulative computation and intermediate data size. Moving from layer 112 to 271 requires $1468.8\,\mathrm{GFLOPs}$ of additional on-board computation for a $402.6\,\mathrm{Mbit}$ reduction in data volume ($50.0\%$). In contrast, moving from layer 271 to 327 requires only $470.3\,\mathrm{GFLOPs}$ for a further $402.6\,\mathrm{Mbit}$ reduction ($99.99\%$ relative to layer 271). Layer 271 therefore has low marginal utility, with a computational cost per bit saved of $3.6 \frac{GFLOPs}{Mbit}$ in the 112--271 segment compared with $1.2 \frac{GFLOPs}{Mbit}$ in the 271--327 segment. Directly moving from layer 112 to 327 requires $1939.2\,\mathrm{GFLOPs}$ while reducing data by $99.997\%$.

\begin{figure}[h]
    \centering
    \begin{tikzpicture}
    \begin{axis}[
        width=\columnwidth,
        height=3.5cm,
        xlabel={Cumulative computation $W(l)$ (GFLOPs)},
        ylabel={Intermediate data $D(l)$ (Mbit)},
        xlabel style={font=\scriptsize},
        ylabel style={font=\scriptsize},
        xmin=-90, xmax=3250,
        ymin=-90, ymax=1760,
        xtick={0,1000,2000,3000},
        xticklabel style={font=\scriptsize},
        yticklabel style={font=\scriptsize},
        xmajorgrids=true,
        ymajorgrids=true,
        grid style={line width=.1pt, draw=gray!30},
        legend style={
            at={(0.5,1.12)},
            anchor=south,
            legend columns=2,
            font=\scriptsize,
            /tikz/every even column/.style={column sep=5pt},
            draw=none
        },
        legend image post style={only marks, mark size=2.5pt, mark options={solid}},
    ]

    \addplot[black, mark=x, mark options={solid, fill=black}, thick, dashed] coordinates {
        (0, 1610.61)
        (1092.00, 805.31)
        (2560.90, 402.65)
        (3031.21, 0.02)
    };
    \addlegendentry{DNNSplit}

    \addplot[orange, mark=*, mark options={solid, fill=orange}, thick] coordinates {
        (0, 1610.61)
        (1092.00, 805.31)
        (3031.21, 0.02)
    };
    \addlegendentry{ANODA}

    \node[font=\scriptsize, anchor=south] at (axis cs:2560.90, 402.65) {layer 271 (pruned)};

    \end{axis}
    \end{tikzpicture}
    \caption{Candidate split points for Swin-V2-T in the computation-communication plane. ANODA retains the lower-convex-hull layers, pruning those above the hull.}
    \label{fig:swin_v2_t_comparison}
\end{figure}
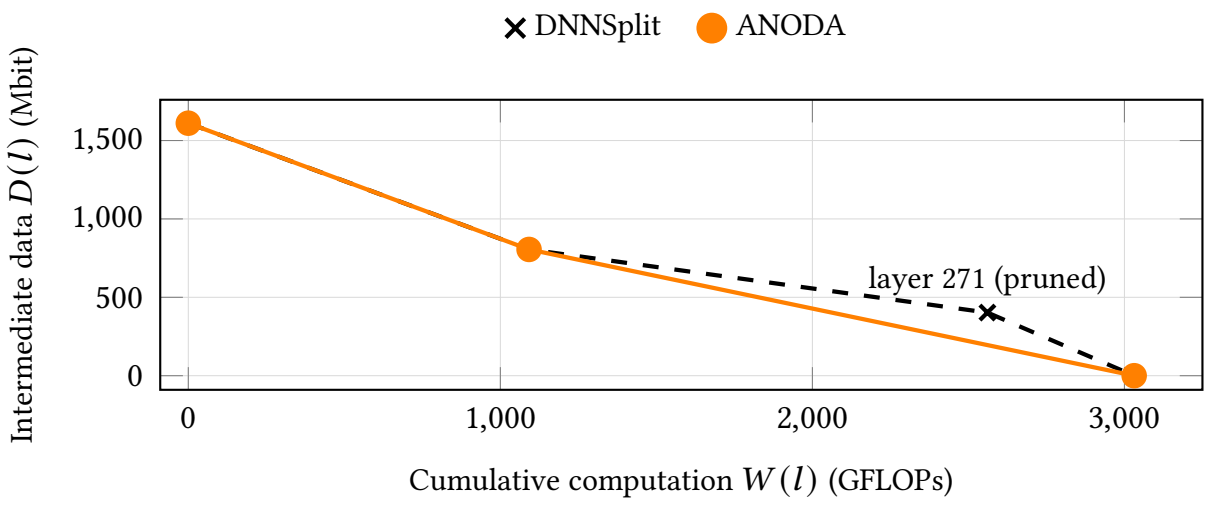

Figure~\ref{fig:swin_v2_t_comparison} presents these points in the computation and communication cost space. ANODA retains layers on the lower convex hull and prunes layer 271, which sits above the hull segment between layers 112 and 327. The slopes between retained layers express the reduction in transmitted data per unit of additional on-board computation, with steeper downward segments indicating greater marginal utility. Layer 271 lies off this frontier and offers low marginal utility for its computation cost, yet DNNSplit retains it.

\begin{figure}[h]
    \centering
    \begin{subfigure}[t]{0.49\columnwidth}
    \centering
    \hspace*{-8mm}
    \begin{tikzpicture}[baseline=(current bounding box.north)]
    \begin{axis}[
        width=\columnwidth,
        height=3.5cm,
        ylabel={Identified Split Points},
        ylabel style={font=\scriptsize},
        xmin=-0.5, xmax=4.5,
        ymin=0, ymax=8,
        ytick={0,2,3,4,5,6,7},
        xtick={0,1,2,3,4},
        xticklabels={
            \texttt{squeezenet1\_1},
            \texttt{swin\_v2\_t},
            \texttt{efficientnet\_b4},
            \texttt{resnet50},
            \texttt{densenet169}
        },
        x tick label style={rotate=35, anchor=east},
        tick label style={font=\scriptsize},
        xmajorgrids=true, ymajorgrids=true,
        grid style={line width=.1pt, draw=gray!30},
        legend style={at={(0.5,1.12)}, anchor=south, legend columns=2, font=\scriptsize,
            /tikz/every even column/.style={column sep=5pt}, draw=none},
    ]
    \node[font=\tiny, gray, anchor=south] at (axis cs:0, 5.12) {0\%};
    \draw[gray, thick] (axis cs:1,4) -- (axis cs:1,3) node[pos=0.5, fill=white, inner sep=0.5pt, font=\tiny] {\textbf{25.0\%}};
    \draw[gray, thick] (axis cs:2,7) -- (axis cs:2,5) node[pos=0.5, fill=white, inner sep=0.5pt, font=\tiny] {\textbf{28.6\%}};
    \draw[gray, thick] (axis cs:3,4) -- (axis cs:3,2) node[pos=0.5, fill=white, inner sep=0.5pt, font=\tiny] {\textbf{50.0\%}};
    \draw[gray, thick] (axis cs:4,6) -- (axis cs:4,3) node[pos=0.5, fill=white, inner sep=0.5pt, font=\tiny] {\textbf{50.0\%}};

    \addplot[black, mark=x, thick, only marks] coordinates { (0,5) (1,4) (2,7) (3,4) (4,6) };
    \addlegendentry{DNNSplit}
    \addplot[orange, mark=*, thick, only marks] coordinates { (0,5) (1,3) (2,5) (3,2) (4,3) };
    \addlegendentry{ANODA}
    \end{axis}
    \end{tikzpicture}
    \label{fig:split_points_models}
    \end{subfigure}
    \hfill
    \begin{subfigure}[t]{0.49\columnwidth}
    \centering
    \hspace*{-4mm}
    \begin{tikzpicture}[baseline=(current bounding box.north)]
    \begin{axis}[
        width=\columnwidth,
        height=3.5cm,
        ylabel={Space Reduction (\%)},
        ylabel style={font=\scriptsize},
        ybar, bar width=5pt,
        ymin=90, ymax=100,
        tick align=inside,
        ytick={90,80,90,100},
        xmin=-0.5, xmax=4.5,
        xtick={0,1,2,3,4},
        xticklabels={
            \texttt{squeezenet1\_1},
            \texttt{swin\_v2\_t},
            \texttt{efficientnet\_b4},
            \texttt{resnet50},
            \texttt{densenet169}
        },
        x tick label style={rotate=35, anchor=east},
        tick label style={font=\scriptsize},
        xmajorgrids=true, ymajorgrids=true,
        grid style={line width=.1pt, draw=gray!30},
        legend style={at={(0.5,1.12)}, anchor=south, legend columns=-1, font=\scriptsize,
            /tikz/every even column/.style={column sep=5pt}, draw=none},
        legend image code/.code={\pgftransformyshift{-0.05cm}
            \draw[#1, fill=#1] (0cm,0cm) rectangle (0.15cm,0.15cm);},
    ]
    \addplot[pattern=north east lines, draw=black, bar shift=-2.5pt] coordinates { (0, 92.54) (1, 98.79) (2, 98.61) (3, 97.73) (4, 99.00) };
    \addplot[fill=orange, draw=orange, bar shift=2.5pt] coordinates { (0, 92.54) (1, 99.09) (2, 99.01) (3, 98.86) (4, 99.50) };
    \legend{DNNSplit, ANODA}
    \end{axis}
    \end{tikzpicture}
    \label{fig:space_reduction_bar}
    \end{subfigure}

    
    \caption{Comparative analysis of search space reduction across DNN architectures: candidate split-point set size (left) and percentage reduction of the total search space (right).}

    \label{fig:combined}
\end{figure}
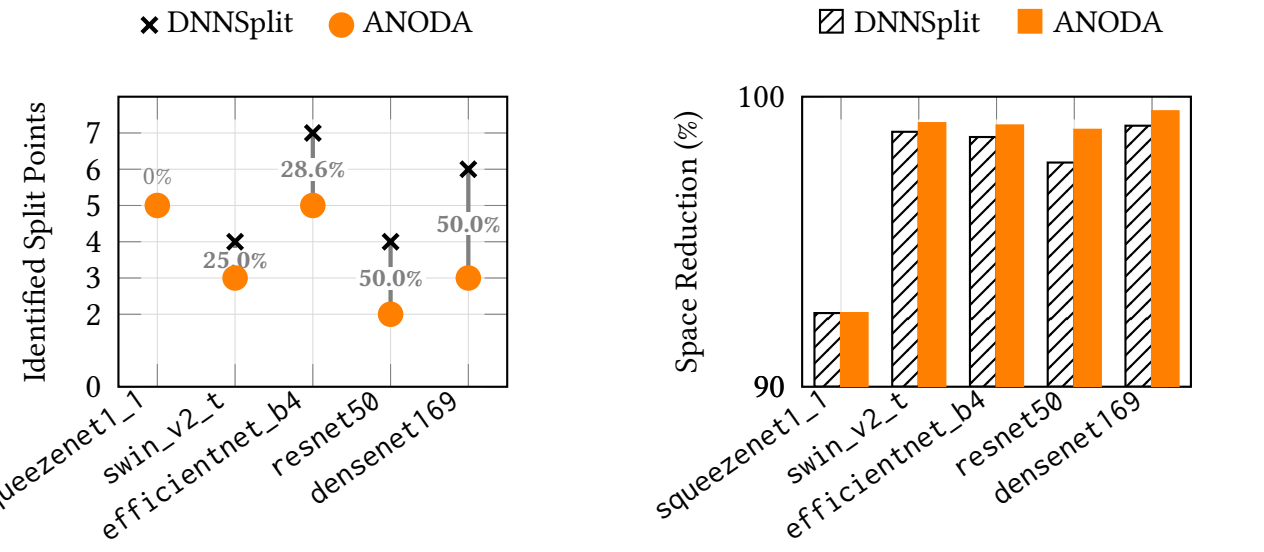

Having established the pruning rationale at the layer level, we next quantify its effect across all evaluated architectures. The left panel of Figure~\ref{fig:combined} shows that ANODA reduces the number of candidate split points by 30.7\% on average relative to DNNSplit. The largest reductions occur for \texttt{resnet50} and \texttt{densenet169}, where ANODA identifies 2 and 3 split points, respectively, compared to 4 and 6 identified by DNNSplit. The right panel of Figure~\ref{fig:combined} normalizes this effect by model depth, showing that both methods reduce the search space by more than 90\% across all evaluated architectures. While both methods compress the search space, ANODA further compacts it by removing dominated split points. {\color{black}To determine how ANODA's candidate-set reduction affects optimization overhead, we measure the per-decision cost of ANELA and the SLICE branch-and-bound optimizer. ANELA is evaluated on the Jetson Orin Nano CPU and SLICE on an Apple M4 Pro CPU to accommodate its higher computational demand.} All reported metrics are averaged over 10 iterations to ensure measurement stability.

Table~\ref{tab:overhead_reduction} reports per-decision latency and energy for three candidate sets. Specifically, the set of all model layers (ALL), the DNNSplit candidate set, and the ANODA candidate set. Relative to DNNSplit, ANODA reduces ANELA per-decision energy by 27.5\% on average across the five models and SLICE energy by 32.2\%. Latency decreases by 25.1\% for ANELA and 30.2\% for SLICE on average across the five models. Considering only the four models where ANODA prunes additional candidates beyond DNNSplit, the latency reductions increase to 31.3\% for ANELA and 38.1\% for SLICE. Relative to the all-layer baseline, both pruned sets reduce per-decision cost by more than an order of magnitude for ANELA and by up to four orders of magnitude for SLICE. For lightweight algorithms such as ANELA, the additional gains of ANODA over DNNSplit are modest but measurable. The SLICE measurements provide an upper-pressure case, showing that the reduction has a larger effect when each retained candidate expands a more expensive search.

\begin{table}[h]
\centering
\caption{Impact of ANODA candidate-set pruning versus DNNSplit on per-decision latency and energy consumption.}
\label{tab:overhead_reduction}
\scriptsize
\setlength{\tabcolsep}{3pt}
\begin{tabular}{@{}llrrrrrr@{}}
\toprule
& & \multicolumn{3}{c}{\textbf{Latency per decision}}
& \multicolumn{3}{c}{\textbf{Energy per decision}} \\
\cmidrule(lr){3-5}\cmidrule(lr){6-8}
\bfseries Model & \bfseries Approach
  & ALL & DNNSplit &  ANODA
  & ALL & DNNSplit &  ANODA \\
\midrule
\multirow{2}{*}{\texttt{squeezenet1\_1}}
  & ANELA & 227.0\,$\mu$s & 20.4\,$\mu$s & 20.4\,$\mu$s & 908\,$\mu$J & 75\,$\mu$J & 75\,$\mu$J \\
  & SLICE & 14.3\,s & 223\,ms & 226\,ms & 95.8\,J & 869\,mJ & 784\,mJ \\
\midrule
\multirow{2}{*}{\texttt{swin\_v2\_t}}
  & ANELA & 823.3\,$\mu$s & 17.1\,$\mu$s & 13.7\,$\mu$s & 3.31\,mJ & 61\,$\mu$J & 48\,$\mu$J \\
  & SLICE & 177\,s & 174\,ms & 133\,ms & 1.16\,kJ & 671\,mJ & 519\,mJ \\
\midrule
\multirow{2}{*}{\texttt{efficientnet\_b4}}
  & ANELA & 1.68\,ms & 27.1\,$\mu$s & 20.4\,$\mu$s & 6.77\,mJ & 102\,$\mu$J & 75\,$\mu$J \\
  & SLICE & 746\,s & 331\,ms & 221\,ms & 4.57\,kJ & 1.27\,J & 850\,mJ \\
\midrule
\multirow{2}{*}{\texttt{resnet50}}
  & ANELA & 590.1\,$\mu$s & 17.1\,$\mu$s & 10.4\,$\mu$s & 2.37\,mJ & 61\,$\mu$J & 35\,$\mu$J \\
  & SLICE & 92.9\,s & 174\,ms & 98.4\,ms & 611\,J & 674\,mJ & 392\,mJ \\
\midrule
\multirow{2}{*}{\texttt{densenet169}}
  & ANELA & 2\,ms & 23.7\,$\mu$s & 13.7\,$\mu$s & 8.07\,mJ & 88\,$\mu$J & 48\,$\mu$J \\
  & SLICE & 1\,081\,s & 276\,ms & 132\,ms & 6.95\,kJ & 1.12\,J & 515\,mJ \\
\bottomrule
\end{tabular}
\end{table}

\subsubsection{STLT results}
\label{subsubsection:stlt_evaluation}

{\color{black}The complete STLT system is evaluated across 220 five-orbit telemetry segments using the five model-specific nominal workloads defined in Section~\ref{subsection:evaluation_setup}. We compare STLT against TOGC, TOBC, RAND, and the satellite-specific SLICE baseline.}

\begin{figure}[ht]
    \raggedright
    \includegraphics[width=0.90\linewidth]{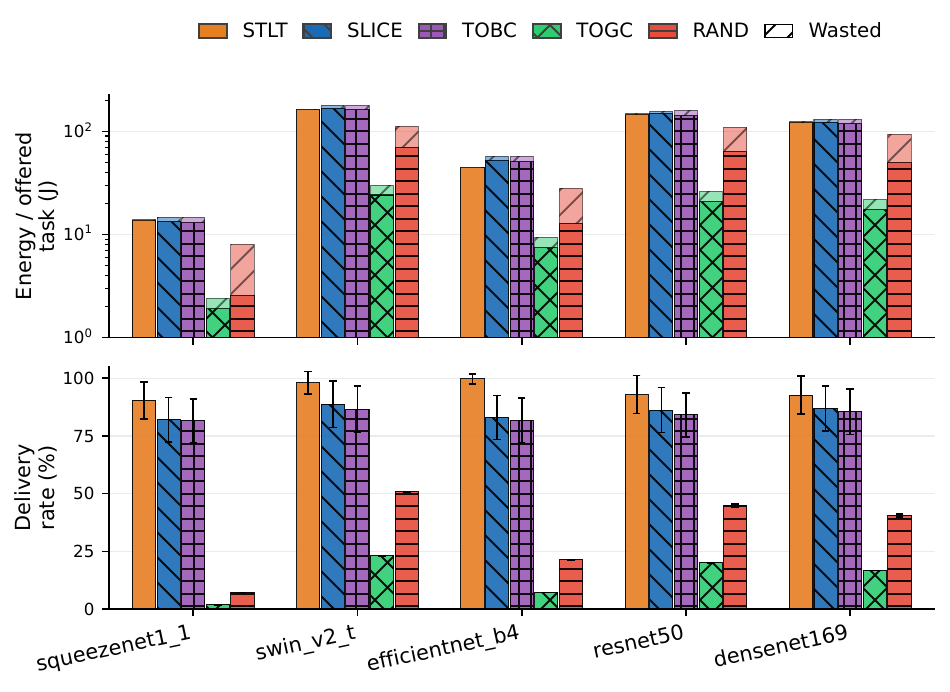}
    \Description{Effective and wasted energy per offered task and mean delivery rate for five DNN models and five policies. Delivery error bars show plus or minus one sample standard deviation across 220 segments. RAND is averaged over five seeds within each segment. Standard-deviation whiskers are not clipped at 100 percent.}
    \caption{{\color{black}Mean EO energy per offered task, separated into effective and wasted energy (top, logarithmic scale), and mean delivery rate (bottom) under the nominal workloads across 220 five-orbit BUPT-1 telemetry segments. Delivery error bars show $\pm$ one sample standard deviation across segments.}}
    \label{fig:adaptive_energy}
\end{figure}

{\color{black}Figure~\ref{fig:adaptive_energy} reports the mean EO energy per offered task and the mean delivery rate, with delivery error bars showing one standard deviation across the 220 segments. RAND is evaluated over five seeds per segment, with results averaged across seeds. Effective energy comprises computation and communication energy spent on delivered tasks, whereas wasted energy comprises the corresponding expenditure on tasks that remain undelivered. Across all models and segments, STLT delivers 256,778 of 277,200 offered tasks, compared with 230,426 for SLICE. This corresponds to 26,352 additional deliveries, an increase of 11.4\%. STLT improves delivery across all five models, with gains ranging from 6.7\% for DenseNet169 to 20.1\% for EfficientNet-B4, while exhibiting lower delivery variability than SLICE and TOBC. For EfficientNet-B4, STLT achieves a mean delivery rate of 99.7\% with a standard deviation of 2.1 percentage points, compared with 83.0\% and 9.5 percentage points for SLICE. STLT consumes 11.84~MJ of EO energy compared with 12.98~MJ for SLICE, saving 1.14~MJ, or 8.8\%, while delivering more tasks. Energy consumption is lower across all five models, with the largest reduction occurring for EfficientNet-B4, at 44.4~J per offered task compared with 56.8~J for SLICE. For this model, SLICE selects full on-board inference at layer~502, whereas ANELA selects between the intermediate split at layer~164 and full inference according to the available EO energy, updated through telemetry replay, and the remaining downlink capacity. STLT also outperforms TOBC, delivering 28,681 additional tasks, an increase of 12.6\%, while saving 1.18~MJ of EO energy, or 9.1\%. TOBC exhibits similar energy consumption and delivery variability to SLICE because both use full on-board inference under the nominal configurations. However, TOBC attempts every task unconditionally, whereas SLICE starts tasks only when their planned computation is affordable, achieving slightly higher delivery with less wasted energy. STLT spends 0.16~MJ on tasks that remain undelivered, compared with 0.91~MJ for SLICE and 1.15~MJ for TOBC, corresponding to 1.3\%, 7.0\%, and 8.9\% of their respective total EO consumption. Although TOGC and RAND consume less energy per offered task and exhibit little delivery variability, both sustain substantially lower delivery rates.}

\begin{figure*}[t]
    \centering
    \includegraphics[width=0.85\textwidth]{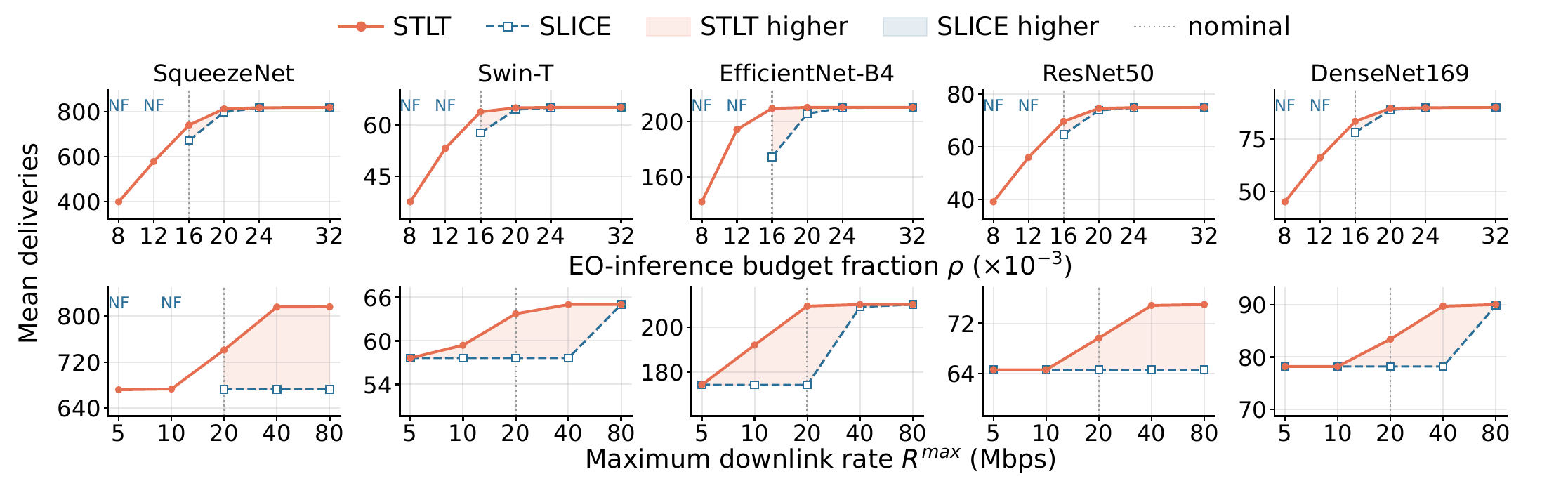}
    \Description{Sensitivity of mean delivered tasks per five-orbit segment across five DNN models and 220 telemetry segments. The top row varies the EO-inference budget fraction from 0.008 to 0.032 at a fixed 20 Mbps downlink rate. The bottom row varies the maximum downlink rate from 5 to 80 Mbps at a fixed budget fraction of 0.016. Workloads remain unchanged and SLICE is replanned at each setting. Transmission power remains fixed at 1.3 W. Dotted lines mark the nominal configuration. Shading indicates differences between policy means, not variability. NF denotes an infeasible complete-workload SLICE plan, not zero delivery.}
    
    
    \caption{{\color{black}Sensitivity of mean deliveries per five-orbit segment to the EO-inference budget fraction $\rho$ and maximum downlink rate $R^{max}$ across 220 segments. The top row varies $\rho$ from 0.008 to 0.032 at $R^{max}=20$~Mbps, while the bottom row varies $R^{max}$ from 5 to 80~Mbps at $\rho=0.016$. Workloads remain fixed, while SLICE is replanned for each setting. Dotted lines mark the nominal configuration, and shading shows policy differences. NF means no feasible SLICE plan, not zero delivery.}}
    
    \label{fig:sensitivity_facets}
\end{figure*}

\begin{table}[h]
    \color{black}
    \centering
\caption{Task outcomes for STLT and SLICE under the nominal workloads. Percentages are relative to offered tasks. Started denotes STLT admissions or SLICE computation starts, while started but not delivered denotes started tasks that fail to reach the ground.}
    \label{tab:task_admission}
    \scriptsize
    \setlength{\tabcolsep}{2.5pt}
    \renewcommand{\arraystretch}{1.08}
    \resizebox{\columnwidth}{!}{%
    \begin{tabular}{@{}llrrrr@{}}
        \toprule
        \multirow{2}{*}{\textbf{Model}} &
        \multirow{2}{*}{\textbf{Policy}} &
        \multicolumn{4}{c}{\textbf{Task outcomes}} \\
        \cmidrule(l){3-6}
        & & \textbf{Offered} & \textbf{Started} & \textbf{Delivered} &
        \shortstack{\textbf{Started but}\\\textbf{not delivered}} \\
        \midrule
        \multirow{2}{*}{SqueezeNet1.1}
            & STLT  & \multirow{2}{*}{180,400 (100\%)} & 166,803 (92.5\%) & 163,039 (90.4\%) & 3,764 (2.1\%) \\
            & SLICE &                                  & 162,760 (90.2\%) & 147,999 (82.0\%) & 14,761 (8.2\%) \\
        \addlinespace[2pt]
        \multirow{2}{*}{Swin-V2-T}
            & STLT  & \multirow{2}{*}{14,300 (100\%)} & 14,110 (98.7\%) & 14,019 (98.0\%) & 91 (0.6\%) \\
            & SLICE &                                 & 13,421 (93.9\%) & 12,679 (88.7\%) & 742 (5.2\%) \\
        \addlinespace[2pt]
        \multirow{2}{*}{EfficientNet-B4}
            & STLT  & \multirow{2}{*}{46,200 (100\%)} & 46,102 (99.8\%) & 46,046 (99.7\%) & 56 (0.1\%) \\
            & SLICE &                                 & 41,730 (90.3\%) & 38,328 (83.0\%) & 3,402 (7.4\%) \\
        \addlinespace[2pt]
        \multirow{2}{*}{ResNet50}
            & STLT  & \multirow{2}{*}{16,500 (100\%)} & 15,737 (95.4\%) & 15,328 (92.9\%) & 409 (2.5\%) \\
            & SLICE &                                 & 15,122 (91.6\%) & 14,219 (86.2\%) & 903 (5.5\%) \\
        \addlinespace[2pt]
        \multirow{2}{*}{DenseNet169}
            & STLT  & \multirow{2}{*}{19,800 (100\%)} & 18,784 (94.9\%) & 18,346 (92.7\%) & 438 (2.2\%) \\
            & SLICE &                                 & 18,337 (92.6\%) & 17,201 (86.9\%) & 1,136 (5.7\%) \\
        \bottomrule
    \end{tabular}
    }
\end{table}

{\color{black}
Table~\ref{tab:task_admission} shows that STLT rejects 5.7\% of offered tasks, while SLICE leaves 9.3\% unstarted because it cannot afford their planned computation. Of the tasks admitted or started, 1.8\% remain undelivered under STLT, compared with 8.3\% under SLICE. These losses can occur when telemetry-driven battery depletion reduces the EO allocation after admission. Reserving transmission energy does not protect that allocation, and queued outputs expire if they cannot be transmitted before the contact ends. TOBC delivers 82.3\% of offered tasks by completing inference on board, limiting communication demand. In contrast, TOGC's raw transmissions saturate the downlink, limiting delivery to 6.0\%, while RAND's feasibility-unaware split choices achieve 16.4\%. ANELA selects the unconstrained minimum, defined as the candidate with the lowest modeled computation-plus-communication energy before feasibility filtering, for 39.3\% of admissions. For the remaining 60.7\%, it selects the lowest-energy alternative passing its feasibility checks. These outcomes illustrate how the changing battery state, remaining EO budget, committed downlink volume, and time to contact can alter split feasibility across task arrivals despite fixed $W(l)$ and $D(l)$ model profiles.
}

{\color{black}Figure~\ref{fig:sensitivity_facets} examines how delivery changes with the EO-inference budget fraction and maximum downlink rate. Each sweep varies one resource parameter while retaining the nominal workloads, computation parameters, and telemetry segments. At $\rho=0.008$, STLT delivers between 48.7\% and 67.5\% of offered tasks across the five models, although SLICE cannot construct a feasible complete-workload plan. SLICE remains infeasible at $\rho=0.012$, whereas STLT continues to admit tasks individually according to the available resources. As the EO allocation increases, the delivery gap narrows, and both policies deliver the complete workload at $\rho=0.032$. Increasing the downlink rate has a different effect because it expands contact capacity and reduces transmission energy per bit at fixed transmission power. At 5~Mbps, STLT and SLICE achieve equal delivery for the four models where SLICE remains feasible, while SLICE cannot construct a feasible plan for SqueezeNet1.1. STLT gains as additional transmission capacity becomes available. SLICE also benefits when replanning selects a different split, reaching complete delivery for Swin-V2-T and EfficientNet-B4 at 80~Mbps and 99.8\% for DenseNet169. For SqueezeNet1.1 and ResNet50, however, SLICE retains full on-board inference and delivers 82.0\% and 86.2\% of offered tasks, compared with 99.5\% and 100\% for STLT. These results show that additional downlink capacity does not necessarily eliminate the benefit of per-task split selection.}



\begin{figure}[h]
    \centering

    \begin{subfigure}[t]{0.49\columnwidth}
        \centering
        \begin{tikzpicture}[baseline=(current bounding box.north)]
        \begin{axis}[
            width=\columnwidth,
            height=3.5cm,
            xmode=log,
            log basis x=10,
            xlabel={Search space $|L^*|$},
            ylabel={Time per decision (ms)},
            xlabel style={font=\scriptsize},
            ylabel style={font=\scriptsize},
            tick label style={font=\scriptsize},
            xmin=4,
            xmax=1200,
            ymajorgrids=true,
            xmajorgrids=true,
            grid style={line width=.1pt, draw=gray!30}
        ]

        \addplot[
            black,
            mark=*,
            thick
        ] coordinates {
            (5,    0.0188)
            (10,   0.0353)
            (20,   0.0677)
            (100,  0.332)
            (500,  1.67)
            (1000, 3.35)
        };

        \end{axis}
        \end{tikzpicture}
        \label{fig:algo_latency}
    \end{subfigure}
    \hfill
    \begin{subfigure}[t]{0.49\columnwidth}
        \centering
        \hspace*{-4mm}
        \begin{tikzpicture}[baseline=(current bounding box.north)]
        \begin{axis}[
            width=\columnwidth,
            height=3.5cm,
            xmode=log,
            log basis x=10,
            xlabel={Search space $|L^*|$},
            ylabel={Energy per decision (mJ)},
            xlabel style={font=\scriptsize},
            ylabel style={font=\scriptsize},
            tick label style={font=\scriptsize},
            xmin=4,
            xmax=1200,
            ymajorgrids=true,
            xmajorgrids=true,
            grid style={line width=.1pt, draw=gray!30}
        ]

        \addplot[
            black,
            mark=*,
            thick
        ] coordinates {
            (5,    0.0757)
            (10,   0.143)
            (20,   0.274)
            (100,  1.35)
            (500,  6.85)
            (1000, 13.7)
        };

        \end{axis}
        \end{tikzpicture}
        \label{fig:algo_energy}
    \end{subfigure}

\caption{ANELA overhead as a function of search-space size, showing average decision latency (left) and energy (right).}
    \label{fig:algo_overhead}
\end{figure}
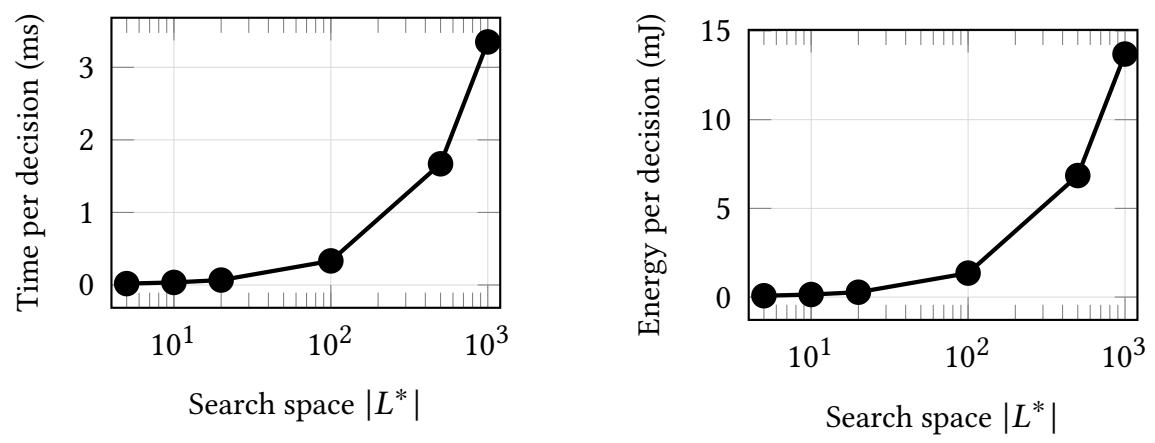

{\color{black}Finally, STLT's per-task adaptation is only practical if ANELA remains lightweight enough to execute before every inference task. Figure~\ref{fig:algo_overhead} characterizes its on-board decision cost as the search space grows. For each size, we execute ANELA on the Jetson Orin Nano CPU and average latency and energy over five measurement runs. Both metrics scale approximately linearly with candidate-set size. Even at $|L^*|=1000$ layers, one decision requires only 3.35~ms and 13.7~mJ. The ANODA candidate sets used in the evaluation contain only two to five split points and therefore remain near the lowest end of this range. Across the five model configurations reported in Table~\ref{tab:overhead_reduction}, ANELA averages 0.0157~ms and 0.056~mJ per decision. These results show that ANODA keeps per-task admission and split-selection overhead suitable for on-board execution.}

\vspace{-0.5cm}
\subsection{Discussion and Limitations}\label{subsec:discussion-limitations}

{\color{black}STLT combines ANODA's offline candidate pruning with ANELA's lightweight on-board admission and split selection. ANODA's lower-hull criterion removes dominated computation-communication trade-offs retained by DNNSplit, producing a candidate set that is never larger than DNNSplit's and is 30.7\% smaller on average across the evaluated architectures. This compact set lets ANELA revise admission and split-point decisions at each task arrival as the battery state and committed downlink volume evolve. Compared with SLICE, STLT delivers 26,352 more tasks while saving 1.14~MJ of EO energy across the nominal workloads and all 220 segments.}

{\color{black}Several modeling assumptions bound these findings. The analytical computation-energy model represents CPU execution. GPUs and other accelerators can be supported through offline measurements of cumulative split-point energy, which can replace the FLOP-based costs used by ANODA without changing ANELA. Sensitivity to errors in these estimates is not evaluated. The communication model assumes a fixed link and does not capture protocol overhead, error correction, or retransmission dynamics. Fixed expected overheads can be represented through effective values of $D(l)$, $\beta$, and $R^{max}$, while nonlinear effects require extending the model and candidate analysis. The solar model assumes active sun-tracking panels with an ideal incidence angle ($\theta \approx 0$), supporting predictable solar input under known orbital illumination. The evaluation uses the recorded solar-generation profile as the forecast of future solar input. Sensitivity to forecast errors is not evaluated, and uncertainty margins could strengthen feasibility decisions.} {\color{black}Several operational boundaries also remain. ANELA runs after image capture, so rejection avoids only the DNN computation and communication energy modeled by STLT. The evaluation excludes energy for sensing, preprocessing, metadata handling, buffering, and storage, as well as the application cost of missed observations. It reports aggregate admission outcomes but does not examine the temporal clustering of rejections or the resulting coverage gaps. The evaluated design assumes sufficient capacity to buffer the selected representation until the upcoming communication window. Feature compression can reduce $D(l)$ but adds on-board computation, so ANODA must account for both effects when deriving the candidate set. STLT targets one-cycle delivery without a cross-cycle downlink backlog and assumes that $\Delta$ exceeds on-board inference time, thereby excluding task queues. Queue-aware scheduling could support deferral, retry, and higher-rate sensing. The current design focuses on satellite-ground communication. An ISL extension could complement queue-aware scheduling by redirecting tasks or intermediate data to satellites with available computation capacity or ground-station contacts.}

\vspace{-0.4cm}
\section{Related Work}
\label{section:related_work}

Energy-efficient satellite-to-ground co-inference in the Edge-Cloud-Space continuum presents two challenges, partitioning neural networks to balance computation and communication costs and adapting these decisions to dynamic satellite constraints. The former is well studied in terrestrial edge-cloud settings and forms the basis of model splitting, while the latter remains an emerging challenge in satellite computing. We review both areas, highlighting limitations that motivate STLT's design.

Model splitting partitions neural network inference between resource-constrained edge devices and remote servers at layer boundaries, with the edge executing initial layers and transmitting intermediate features for remote completion. Neurosurgeon~\cite{10.1145/3037697.3037698} introduced layer-level partitioning that exhaustively evaluates split points to optimize end-to-end latency and energy, but relies on static profiling and sequential architectures. JointDNN~\cite{8871124} extended this approach to Directed Acyclic Graph architectures by formulating partitioning as a shortest-path problem solved through Integer Linear Programming, but requires $O(N^2)$ profiling, creating substantial overhead as model depth increases. DADS~\cite{DADS} introduced adaptive partitioning through min-cut formulations for light-load latency and heavy-load throughput optimization. As modern DNNs grow to hundreds of layers, exhaustive search becomes computationally intractable. EdgePipe~\cite{EDGEPIPE} and DNNPipe~\cite{DNNPIPE} employ dynamic programming and pruning to reduce the search space, yet their worst-case complexity remains prohibitive for real-time adaptation. {\color{black}DNNSplit~\cite{DNNSplit} reduces search complexity by selecting candidates from decreases in intermediate data size, reporting over $100\times$ lower partitioning complexity for ResNet-101. However, this criterion does not account for the cumulative on-board computation required to reach each candidate. Existing model-splitting methods largely target terrestrial edge-cloud settings and often rely on optimization procedures unsuitable for per-task execution on resource-constrained satellites. STLT adapts model splitting to satellite-ground operation by jointly evaluating cumulative computation and intermediate data size offline, retaining lower-hull layers that can minimize the linear split-point energy objective. The resulting compact candidate set supports lightweight on-board admission and split selection based on the current satellite state.}

The proliferation of LEO satellites with commodity hardware has motivated on-board computation as an alternative to ``bent-pipe'' architectures. Orbital Edge Computing~\cite{10.1145/3373376.3378473} showed that nanosatellite constellations can act as distributed compute platforms while reducing downlink requirements. FOOL~\cite{FOOL} addresses the downlink bottleneck through neural feature compression, transmitting compact representations instead of raw imagery. SatCooper~\cite{zhang2025satcooper} uses multi-exit DNNs to terminate inference once a confidence threshold is met, further reducing transmission.

Recent work optimizes satellite-ground co-inference under energy and latency constraints. SLICE~\cite{SLICE} schedules splits layer-wise using branch-and-bound with Lagrange multipliers to minimize satellite energy, reporting 96\% average savings over raw transmission. Other approaches~\cite{MDP} formulate partitioning as a Markov Decision Process and use Deep Reinforcement Learning to select partition points from battery level and channel state. AOSR~\cite{AOSR} combines DRL with Karush-Kuhn-Tucker conditions to jointly optimize split ratios and resource allocation, including Dynamic Voltage and Frequency Scaling. These adaptive frameworks rely on extensive layer-specific profiling or historical telemetry, difficult to maintain under short communication windows and rapidly varying LEO links, while DRL variants add training and policy-update overhead impractical on-board. {\color{black}STLT instead selects splits using current satellite state, without historical telemetry or learned policies.}
\vspace{-0.2cm}

\section{Conclusion and Future Work}
\label{section:conclusion}

LEO satellites with COTS hardware enable on-board DNN inference, but energy constraints demand careful management of the computation-communication trade-off. This work presented STLT, {\color{revisiongreen}a two-stage system that coordinates energy-aware distributed DNN inference between LEO satellites and ground stations by combining offline model analysis with lightweight on-board admission and split selection.} ANODA applies convex-hull analysis to extract a candidate split set 30.7\% smaller than DNNSplit on average, and the lightweight on-board ANELA {\color{revisiongreen}performs admission and split selection for each inference task using the current satellite state.} {\color{black}Across the evaluated nominal workloads, STLT delivers up to 20.1\% more inference tasks than SLICE, with an aggregate delivery improvement of 11.4\% and 8.8\% lower total EO energy consumption}, while pruning lowers per-decision latency and energy by 25.1\% and 27.5\% on average at $0.0157\,\mathrm{ms}$ and $0.056\,\mathrm{mJ}$ per decision. Future work will incorporate ISL to distribute inference across constellations, and extend STLT to the full Edge-Cloud-Space continuum with edge nodes for multi-tier distributed inference.

\begin{acks}
This work was partly funded by the European Union under the Horizon Europe programme through the SNS JU (Grant Agreement No. 101192912, NexaSphere). Views expressed are those of the authors and do not necessarily reflect those of the EU or SNS JU.
\end{acks}

\bibliographystyle{ACM-Reference-Format}
\bibliography{references}

\end{document}